\documentclass[preprintnumbers,aps,prb,twocolumn]{revtex4}

\usepackage{graphicx}% Include figure files
\usepackage{dcolumn} % Align table columns on decimal point
\usepackage{amsmath}
\usepackage{amssymb}
\usepackage{bm}      % bold math
\usepackage{graphicx}
\usepackage{color}
\usepackage{epsfig,psfrag,subfigure,subeqnarray}
\usepackage{bbm}
\usepackage{enumitem}
\def\lan{\langle}
\def\ran{\rangle}
\def\va{\varepsilon}

\def\vk{{\bf k}}

\def\vr{{\bf r}}

\def\vq{{\bf q}}
\def\vp{{\bf p}}

\newcommand{\bd}{\begin{equation}}
\newcommand{\ed}{\end{equation}}
\newcommand{\be}{\begin{equation}}
\newcommand{\ee}{\end{equation}}
\newcommand{\bt}{\begin{split}}
\newcommand{\et}{\end{split}}

\newcommand{\bn}{\begin{align}}
\newcommand{\en}{\end{align}}

\newcommand{\bea}{\begin{eqnarray}}
\newcommand{\eea}{\end{eqnarray}}
\newcommand{\ba}{\begin{array}}
\newcommand{\ea}{\end{array}}
\newcommand{\nn}{\nonumber}

\begin{document}
\title {Correlation Energy for Elementary Bosons: Physics of the Singularity}

\author{Shiue-Yuan Shiau$^{1}$, Monique Combescot$^2$  and Yia-Chung Chang$^{3,1}$}
\email{yiachang@gate.sinica.edu.tw}
\affiliation{$^1$ Department of Physics, National Cheng Kung University, Tainan, 701 Taiwan}
\affiliation{$^2$Institut des NanoSciences de Paris, Universit\'e Pierre et Marie Curie, CNRS, 4 place Jussieu, 75005 Paris}
\affiliation{$^3$Research Center for Applied Sciences, Academia Sinica, Taipei, 115 Taiwan}
\date{\today}

\begin{abstract}
We propose a compact perturbative approach that reveals the physical origin of the singularity occurring in the density dependence of correlation energy:  like fermions, elementary bosons have a singular correlation energy which comes from the accumulation, through Feynman ``bubble" diagrams, of the same non-zero momentum transfer excitations from the free particle ground state, that is, the Fermi sea for fermions and the Bose-Einstein condensate for bosons. This understanding paves the way toward deriving the correlation energy of composite bosons like atomic dimers and semiconductor excitons, by suggesting Shiva diagrams that have similarity with Feynman ``bubble" diagrams, the previous elementary boson approaches, which hide this physics, being difficult to do so.

%Using a compact perturbative approach, we show that, although bosons and fermions have different quantum nature and different correlation energy density dependence, ($n^{3/2}$) and ($\ln n$), these singular behaviors just come from the same physics
%as the one for the singular ($\ln n$) dependence of the electron correlation energy
%: the piling up, through ``bubble" Feynman diagrams, of same $\vq\not= \bf0$ excitations from (different) free particle ground state,
%\textit{ i.e.,}
 %the Bose-Einstein condensate and the Fermi sea. This understanding paves the way toward deriving the correlation energy of composite bosons made of fermion pairs by suggesting the appropriate set of Shiva diagrams to sum up.
\end{abstract}

\maketitle

\section{Introduction}
The energy of $N$ interacting elementary bosons has been studied in the late 50's by Brueckner and Sawada\cite{BruecknerPR1957}, and by Lee, Huang and Yang\cite{LY1957,LHY1957} through a mean-field procedure which transforms the two-body Hamiltonian into a quadratic operator easy to diagonalize by using a Bogoliubov-like transformation. Their most striking result is that the $N$-boson correlation energy is singular with a dependence on density $n=N/L^3$ in $n^{3/2}$ instead of $n^2$. More precisely, the
ground-state energy of $N$ interacting bosons reads as
\be
\frac{\mathcal{E}_N}{N}\,\simeq \,\frac{2\pi}{ma^2}\left[na^3+\frac{128 }{15\sqrt{\pi}}{(n a^3)}^{3/2}\right]\label{eq:MFenergy}
\ee
where $m$ is the boson mass and $a$ is the scattering length of the boson-boson potential at hand. Actually, the above result with the scattering length appearing also in the $n^{3/2}$ term has been obtained by Lee, Huang and Yang, but not by Brueckner and Sawada. The reason is that the former authors use a pseudo-potential which reads in terms of $a$; so, they do not have to bother about generating the scattering length in the correlation term. The drawback of these two previous approaches is that they rely on a mean-field approximation which completely hides the physics of the correlation energy singularity. A more elaborate field-theory procedure has been later proposed\cite{Hugenholtz1959}. It provides a better control of the performed approximations, but through a more complicated procedure that still hides the physical origin of the singularity.\

%The correlation energy\cite{Gellmann1957} of $N$ electrons is also known to be singular with a density dependence ($\ln n$) instead of $n^0$, this result being often written as ($\ln r_s$) where $r_s$ is the inter-electron distance in Bohr radius unit.
% $a_B$, defined for $N$ electrons in size L sample through
%\be
%N(\frac{4}{3}\pi r_sa_B)^3=L^3
%\ee

The purpose of this paper is to reveal that, in spite of the different quantum nature of the particles and the different singular dependence in the correlation energy of bosons and fermions\cite{Gellmann1957}, namely $(n^{3/2})$ and ($\ln n$), the physics producing these singular dependences is just the same: the accumulation of the same non-zero momentum transfer excitations from the free particle ground state, that is, the $\textbf{k}=\textbf{0}$ boson condensate or the $0\leqslant |\textbf{k}|\leqslant k_F$ Fermi sea. This understanding paves the way toward deriving the correlation energy of $N$ composite bosons made of fermion pairs by suggesting the appropriate set of Shiva diagrams\cite{moniqPhysRep} that have similarity with the Feynman ``bubble" diagrams leading to the singular correlation energy of elementary bosons. Composite bosons of major present interest are atomic dimers\cite{Petrov2004,Alzetto2013,Leyronas2007} made of different species of cold fermionic atoms\cite{Pitaevskiibook,Hu2007,Bloch2012}, semiconductor excitons\cite{Yoshioka2011,High2012,Alloing2014} made of electrons and holes, and polaritons\cite{Deng2002,Kasprzak2006,Balili2007,Baumberg2008} which are linear combination of photons and excitons. Cold atoms\cite{Stamper1999,Ernst2010} and polaritons\cite{Utsun2008} have been used as a testbed for low-energy (Bogoliubov) excitations in the mean-field  framework of elementary bosons. Whether composite bosons defy this mean-field description because of their composite nature remains under debate in spite of the fact that, because of the fermion indistinguishability, one cannot construct an effective potential between composite bosons that is valid beyond first order in interaction\cite{moniqPhysRep}. As a direct consequence, elementary boson approaches based on the existence of a boson-boson potential cannot be duplicated for composite boson systems.\

 To show the analogy between the correlation energies of elementary bosons and elementary fermions, we propose a compact perturbative approach to the energy of $N$ quantum particles that allows catching the physics of the singular interaction processes. Once selected and summed up, these singular processes lead to the $N$-boson energy given in Eq.~(\ref{eq:MFenergy}). Unlike previous methods, the perturbative approach that we here propose can be directly extended to composite bosons which interact not only through the fermion-fermion interactions between their fermionic components, but also through fermion exchanges.

The paper is organized as follow:

$\bullet$ We first give some general arguments for understanding what should be and what really is the density dependence of the correlation energies for $N$ elementary bosons and $N$ elementary fermions, in order to understand why these density dependences end up being different, albeit produced by the same physical processes.

$\bullet$ Next, we propose a compact perturbative approach to derive the $N$-boson energy, which allows catching the physics of its various terms in a transparent way. We also provide the key commutators which enable us to calculate these terms easily.

$\bullet$ We then recover the $N$-boson energy given in Eq.~(\ref{eq:MFenergy}) through the explicit summations of the ladder diagrams associated with the scattering length, and of the bubble diagrams associated with the correlation energy singularity. We also discuss the required cancellation of overextensive contributions that come from disconnected diagrams, as standard in perturbative expansion.

$\bullet$ We conclude with the state-of-the-art for composite boson systems and the fundamental problem raised by the Pauli exclusion principle between the fermionic components of composite quantum particles.

 In the supplemental material, we outline the original derivations of the $N$-boson energy proposed by Brueckner and Sawada, and by Lee, Huang and Yang, the physics responsible for the correlation energy singularity being hard to catch from these mean-field approaches. The supplemental material also contains some heavy diagrammatic parts associated with high-order perturbative expansions, which support our procedure but are not necessary to follow its spirit.

\section{General arguments}
We look for the solution of
\be
(H-\mathcal{E}_N)|\psi_N\ran=0\, , \label{eq:EN1}
\ee
 where $H=H_0+V$  is the Hamiltonian of the quantum particles at hand. The free part is given by
 \be
 H_0=\sum_\vk \va_\vk c^\dag_\vk c_\vk\label{def:H0}
 \ee
 with $\va_\vk=\vk^2/2m$ while the interaction part can be written as
 \be
V=\frac{1}{2}\sum_{\vq}v_\vq\sum_{\vk\vk'}c_{\vk+\vq}^\dag c_{\vk'-\vq}^\dag c_{\vk'} c_\vk\ .\label{def:V1}
\ee

The elementary particle operators fulfill
\begin{subeqnarray}
\left[c^\dag_\vk,c^\dag_{\vk'}\right]_\eta&=&c^\dag_\vk c^\dag_{\vk'}+\eta c^\dag_{\vk'} c^\dag_\vk=0\, ,\\
%\left[c^\dag_\vk,c^\dag_\vk' \right]=0\nn\\
%\ee
%\be
\left[c_\vk,c^\dag_{\vk'}\right]_\eta&=&\delta_{\vk,\vk'}\, ,
\end{subeqnarray}
with $\eta=-1$ for bosons, $c^\dag_\vk\equiv b^\dag_\vk$, and $\eta=+1$ for fermions,  $c^\dag_\vk\equiv a^\dag_\vk$. Since $(a^\dag_\vk)^2=0$, two fermions cannot be in the same state; so, the $H_0$ ground state corresponds to a Fermi sea having all $\vk$ states with $0\leqslant |\textbf{k}|\leqslant k_F$ occupied, the number of 3D fermions with up and down spins being related to the Fermi momentum $k_F$ through
\be
N=2\sum_{{\bf0}\leqslant \vk\leqslant \vk_F} 1=\frac{L^3k_F^3}{3\pi^2}\, ,\label{eq:6}
\ee
the extra $2$ coming from spin. By contrast, since $(b^\dag_\vk)^2\neq0$, two bosons can be in the same state. So, the $H_0$ ground state corresponds to the so-called ``condensate" with all bosons in the lowest energy state.
% $\vk=\textbf{0}$.

To understand that the difference in the density dependence of $\mathcal{E}_N$ for bosons and fermions comes from the same physics but within different $H_0$ ground states, let us expand $\mathcal{E}_N$ in terms of the number of particles involved in scattering processes, namely
\be
\mathcal{E}_N=(\cdots)N+(\cdots) N(N-1)+(\cdots) N(N-1)(N-2)+\cdots \, .\label{expansion1}
\ee
The fact that the Fermi sea extension depends on fermion number through Eq.~(\ref{eq:6}) brings density-dependent prefactors into the above expansion, while such case does not exist for bosons. So, the resulting density dependences of $\mathcal{E}_N/N$ for bosons and fermions have to be different, even if they are induced by same physical processes. 

This is already seen from the $N$ term which corresponds to the energy in the absence of interaction. The energy of $N$ free bosons in the $\vk=\textbf{0}$ condensate reduces to $0$ while the energy of $N$ fermions in the Fermi sea scales as $N\va_F\propto N n^{3/2}$.

To analyze the other terms, we use dimensional arguments. As the $N$-particle energy is an extensive quantity, $\mathcal{E}_N/N$ only depends on density. Potential scatterings depend on sample volume as $1/L^3$, while each sum over momentum brings a $L^3$ factor when transformed into an integral. Let us successively consider what these comments imply on the boson and fermion energies.

\textbf{A. Bosons} 
 
$\bullet$  The $N(N-1)$ term of $\mathcal{E}_N$ comes from $(1,2,3,\cdots)$ interactions between two bosons taken among $N$.

 As these two bosons come from the condensate, process involving a single potential scattering can only appear for $\textbf{q}=\textbf{0}$. So, it gives, within a numerical prefactor,
\be
 N(N-1) v_\textbf{0}\propto Nn\, .\label{eq8}
\ee

Processes involving two potential scatterings must contain a $\textbf{q}$ sum to have the proper extensivity, and an energy denominator to be an energy-like quantity. This energy denominator can only be the energy of the boson pair $(\vq,-\vq)$ excited from the condensate, as a result of the $\textbf{q}$ scattering. This gives, within a numerical prefactor, 
\be
 N(N-1) \sum_{\vq_1}\frac{v_{\vq_1}^2}{2\va_{\vq_1} }\propto Nn\, . \label{eq9}
\ee

Processes involving three potential scatterings must contain two $\textbf{q}$ sums, and two energy denominators. And so on.

 This shows that the $Nn$ terms of $\mathcal{E}_N$ come from interaction between two bosons excited out of the condensate through the so-called ``ladder processes", each additional $v_\vq $ being accompanied with a $\vq $ sum and a $2\va_\vq $ denominator, in order to make this additional part dimensionless and sample volume free.

 Such ladder processes physically correspond to the renormalization of the $v_\textbf{0}$ scattering appearing in Eq.~(\ref{eq8}), through the repeated excitations of \textit{one} boson pair from the condensate, this pair changing from ($\vq_1,-\vq _1$) to ($\vq_2,-\vq _2$), and so on. These ladder processes change the bare first-order term of Eq.~(\ref{eq8}) into
\be
 N(N-1) \tilde v_\textbf{0}\propto Nn\, ,\label{eq10}
\ee
the renormalized $\textbf{q}=\textbf{0}$ scattering being commonly written in terms of the scattering length $a$ as
\be
 \tilde v_\textbf{0}=\frac{4\pi a}{mL^3}\, .
\ee

$\bullet$  The $N(N-1)(N-2)$ term of $\mathcal{E}_N$ comes from interaction between three bosons.
To be an energy-like quantity with proper extensivity, this term must contain $m\geqslant2$ interactions, ($m-1$) energy denominators, and ($m-2$) sums over momentum. 

The term with two scatterings does not exist, because, in the absence of $\vq $ sum, it can only contain the $ \vq =\textbf{0}$ process; so, the required energy denominator $2\va_\vq $ would be zero.

The term with three potential scatterings, which must have one $\vq $ sum and two energy denominators, reads, within a numerical prefactor, as
\be
 N(N-1)(N-2) \sum_{\vq_1}\frac{v_{\vq_1}^3}{(2 \va_{\vq_1})^2 }\propto Nn^2\, .\label{Nn^2_01}
\ee

The term with four potential scatterings, which must have two $\vq $ sums and three energy denominators, reads
\be
 N(N-1)(N-2) \sum_{\vq_1}\frac{v_{\vq_1}^2}{(2\va_{\vq_1})^2 } \sum_{\vq_2}\frac{v_{\vq_1-\vq_2}v_{\vq_2}}{2\va_{\vq_2} }\propto Nn^2\, .
\ee
 Although not obvious from dimensional arguments alone, these four potential scatterings split as written above, the $\vq_2$ sum belonging to the ladder processes that  renormalize the  $v_{\vq}$ interaction between two bosons according to 
\be
\tilde v_{\vq_1}=v_{\vq_1}+(\cdot\cdot\cdot)\sum_{\vq_2}v_{\vq_1-\vq_2}\frac{1}{2\va_{\vq_2} }v_{\vq_2}+\cdots\, . \label{eq:ladderprocess}
\ee
Higher-order terms in this potential renormalization come from more than four scatterings between two bosons.

$\bullet$  The $N(N-1)(N-2)(N-3)$ term of $\mathcal{E}_N$ comes from interaction between four bosons.
The energy-like contribution with proper extensivity involves four interactions at least. It reads 
\be
 N(N-1)(N-2)(N-3) \sum_{\vq_1}\frac{v_{\vq_1}^4}{(2\va_{\vq_1})^3 }\propto Nn^3\, .
\ee
 More interactions between four bosons lead to a renormalization of the potential scattering, as in the case of three bosons. And so on.
 
 The $N$-boson correlation energy results from interaction between more than two bosons. It fundamentally reads, within a renormalization of the potential scattering through ladder processes,  as
\bea
&&(\,\cdots)
 N(N-1)(N-2)\sum_{\vq_1}
 \frac{v_{\vq_1}^3}{(2\va_{\vq_1})^2 }\label{diagrams:singular}\\
 &&+(\, \cdots)  N(N-1)(N-2)(N-3)\sum_{\vq_1}\frac{v_{\vq_1}^4}{(2\va_{\vq_1})^3 }+\cdots\, .\nn
\eea

We then note that, for $v_{\vq\rightarrow \bf0}\neq0$, the above $\vq_1$ sums diverge in the small $\vq_1$ limit, with similar divergences occurring for more than four bosons. As standard, the summation of these singular terms overcomes their divergences and produces a finite contribution to the $N$-boson energy with a density dependence somewhat larger than its $Nn^2$ first term. To show that the summation of these singular processes leads to a $n^{3/2}$ dependence requires the knowledge of the numerical prefactors in Eq.~(\ref{diagrams:singular}). These prefactors, which depend on the number of ways these scattering processes appear, are obtained from a precise counting that cannot be done from dimensional arguments only.

\begin{figure}[h!]
\centering
  \includegraphics[trim=2.5cm 5.9cm 3.8cm 20cm,clip,width=3in] {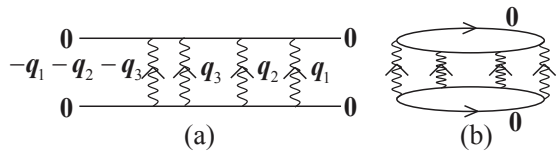}
\caption{\small (a) Ladder process between two bosons from the condensate. (b) Feynman diagram obtained by closing the $\bf0$ lines. }
\label{fig:1}
\end{figure}

We conclude from the above analysis that the repeated interaction between two bosons taken from the $N$-boson condensate produces a dressing of their bare interaction  through a set of ladder processes like the one\cite{diagram} of Fig.~\ref{fig:1}, the resulting linear term in density being given by Eq.~(\ref{eq10}).

\begin{figure}[h!]
\centering
   \includegraphics[trim=2.5cm 3.2cm 2.5cm 16.2cm,clip,width=3.4in] {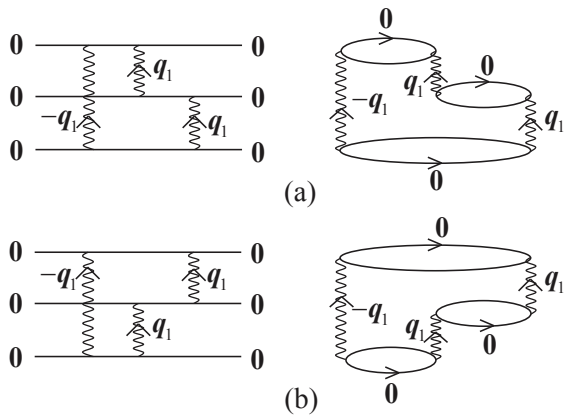}
\caption{\small Bubble processes resulting from three interactions between three bosons $\bf0$ excited from the condensate. They provide equal contributions to the $N$-boson correlation energy.}
\label{fig:2}
\end{figure}

Higher-order terms in density come from interactions between more than two bosons. The most singular ones, given in Eq.~(\ref{diagrams:singular}), correspond to repeatedly exciting and de-exciting the \emph{same} $(\vq_1,-\vq_1)$ pair from the condensate through ``bubble" diagrams like the ones shown in Fig.~\ref{fig:2}.

%\subsection{Fermions}

\textbf{B. Fermions}

$\bullet$  The $N(N-1)$ term of $\mathcal{E}_N$ comes from interaction between two fermions taken from the Fermi sea.

The energy-like term involving a single interaction scales as
\be
 N (N-1)v_{\vk_F}\ \  \propto\ \  Nn^{1/3}\, ,
\ee
%Indeed, since for $v_\vq$ scales as $1/L^3$ as $w/L^3vq
for $v_\vq$ taken as the 3D Coulomb scattering, $v_\vq=4\pi e^2/\epsilon_r L^3 q^2$ for $\vq\neq\bf0$, and $v_{\textbf{0}}=0$, the $\vq=\textbf{0}$ process being eliminated for electrons in a positive ion jellium which ensures the system neutrality \cite{Gellmann1957}. Since only $\vq\neq\textbf{0}$ processes exist, this single-interaction term comes from fermion exchange inside the Fermi sea.
\begin{figure}[h!]
\centering
  \includegraphics[trim=1.5cm 6.1cm 1cm 19cm,clip,width=3.5in] {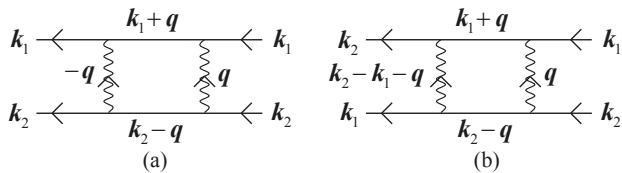}
\caption{\small (a) Direct Coulomb process and (b) exchange Coulomb process between two fermions from the Fermi sea.  }
\label{fig:3extra}
\end{figure}

The energy-like term involving two interactions must have one energy denominator which scales as $\va_{\vk_F}$, and one $\vq$ sum which brings a $N$ factor. So,
it scales as
\be
N (N-1)
 \bigg\{
 N\frac{v^2_{\vk_F}}{\va_{\vk_F}}
 \bigg\}
 \ \ \propto \ \ Nn^{0}\, .\label{2boson2sca}
\ee
With two interactions, there in fact exist direct and exchange processes, as shown in Fig.~\ref{fig:3extra}. The direct process, associated with a two-bubble Feynman diagram, is singular due to the infiniteness of its two $v_\vq$ scatterings in the small $\vq$ limit.

Ladder processes similar to the ones existing between two bosons like Fig.~\ref{fig:1} also exists between two fermions. However, as each interaction comes with a $v_\vq$ scattering which scales as $v_{\vk_F}$, an energy denominator which scales as $\va_{\vk_F}$ , and a $\vq$ sum which leads to a factor $N$, each rung of a ladder process scales as
\be
N\frac{v_{\vk_F}}{\va_{\vk_F}}
\ \ \propto \ \ n^{-1/3}\, ,\label{addfactor}
\ee
 which is small in the large density limit. This is why ladder processes and scattering length are never considered for Coulomb interaction between electrons in the dense limit, while they play a key role for bosons in the dilute limit. Note that this is not so for electron-hole systems due to the poles these ladder processes produce that are associated with exciton formation in the dilute regime.

$\bullet$  The $N(N-1)(N-2)$ term of $\mathcal{E}_N$ comes from interaction between three fermions.
The term with two potential scatterings does not exist for the same reason as that for bosons. The next-order term with three potential scatterings, two energy denominators, and one $\vq$ sum, scales as
\be
 N (N-1)(N-2)
 \bigg\{
 N\frac{v^3_{\vk_F}}{\va^2_{\vk_F}}
 \bigg\}
 %(Lk_F)^3\frac{1}{(k^2_F/2m)^2}v^3_{k_F}
   \ \ \propto\ \   N n^{-1/3}\, .\label{3boson3sca}
\ee

The energy-like term involving four fermions, four scatterings, three energy denominators, and one $\vq$ sum, scales as
\bea
%&&
N (N-1)(N-2)(N-3)
 \bigg\{
 N\frac{v^4_{\vk_F}}{\va^3_{\vk_F}}
 \bigg\}
%(Lk_F)^3\frac{1}{(k^2_F/2m)^3}v^4_{k_F}
 % \nn\\
  %&&
  \ \ \propto \ \ N n^{-2/3}\, ,\label{4boson4sca}
\eea
and so on.\

Again, direct and exchange processes exist for terms involving more than two fermions. Precise calculations\cite{Gellmann1957} show that, much like in the case of bosons, the numerical
%The above equations (\ref{2boson2sca},\ref{3boson3sca},\ref{4boson4sca}) actually belong to the so-called ``bubble diagrams". Their dimensionless
factors of the direct terms diverge in the small momentum transfer limit. When summed up, these divergences transform the $n^0$ dependence of the first term, given in Eq.~(\ref{2boson2sca}), into a somewhat larger $(\ln n)$ dependence. These singular terms correspond to bubble processes, just like the singular terms for bosons that leads to a $n^{3/2}$ dependence for the correlation energy. This understanding is commonly known for fermions but not so for bosons. The unique but major difference between fermions and bosons is that, for fermions, these singular terms are associated with excitations from a Fermi sea, whereas, for bosons, the singular terms are associated with excitations from a condensate with all particles in the $\vk=\textbf{0}$ ground state.\

 Before turning to the procedure we propose to derive the $N$-boson energy given in Eq.~(\ref{eq:MFenergy}), we wish to make an important comment. In the above arguments, we start with the free quantum particle ground state and we perform similar expansion in terms of the number of particles involved in scattering processes. Yet, we end up with a small density expansion in the case of bosons, and with a large density expansion in the case of fermions. The reason is that, in order for the free particle ground state to be a good starting point, this state must not be very much changed by interaction. Such is the case for a very large Fermi sea,\textit{ i.e.}, a very dense fermion system. By contrast, the $N$-boson condensate is extremely narrow in energy by construction; so, to have it not very much changed by interaction, the number of scattering events, \emph{i.e.}, the boson density, must be very small.

\section{Equation fulfilled by the $N$-boson ground-state energy}

We want to solve Eq.~(\ref{eq:EN1}) in the case of bosons. For $N/L^3(=n)$ and $V$ small enough, the $H$ ground state stays close to the $H_0$ ground state, that is the $N$-boson condensate $|0_N\ran=b_{\bf0}^{\dag N}|v\ran$, where $|v\ran$ denotes the vacuum state. We force this $|0_N\ran$ state into the problem by inserting
%We look for the ground state of $N$ interacting elementary bosons, the Hamiltonian being $H=H_0+V$ where $H_0=\sum_\vk \va_\vk b^\dag_\vk b_\vk$ with $\va_\vk=\vk^2/2m$, and
%\be
%V=\sum_{\vq}\frac{v_\vq}{2}\sum_{\vk\vk'}b_{\vk+\vq}^\dag b_{\vk'-\vq}^\dag b_{\vk'} b_\vk\, .\label{def:V}
%\ee
%For weak interaction, we expect the interacting ground state to be close to $|0_N\ran=b_{\bf0}^{\dag N}|v\ran$ where $|v\ran$ denotes the vacuum state. So, to solve\be
%(\mathcal{E}_N-H)|\psi_N\ran=0\, ,\label{eq:EN1}
%\ee

\be
{\rm I}=\frac{|0_N\ran\lan 0_N|}{\lan 0_N|0_N\ran} +P_\perp
\ee
 in front of $|\psi_N\ran$ in Eq.~(\ref{eq:EN1}) and we multiply the resulting equation either by $|0_N\ran$ or by $P_\perp$. As $H_0|0_N\ran=0$, we get
\be
0=\lan 0_N|V-\mathcal{E}_N|0_N\ran \frac{\lan 0_N|\psi_N\ran}{\lan 0_N|0_N\ran}+\lan 0_N| VP_\perp|\psi_N\ran\, \label{eq:EN2}.
\ee
%To get $P_\perp|\psi_N\ran$, we, instead of projecting the resulting equation (\ref{eq:EN1}) on $|0_N\ran$, multiply it by $P_\perp$. 
As $P_\perp H_0|0_N\ran=0$,  multiplication by $P_\perp$ yields
\bea
0&=&P_\perp H_0 P_\perp|\psi_N\ran+P_\perp V |0_N\ran \frac{\lan 0_N|\psi_N\ran}{\lan 0_N|0_N\ran}\nn\\
&&+P_\perp (V-\mathcal{E}_N)P_\perp|\psi_N\ran\, ,
\eea
from which we get
\bea
P_\perp |\psi_N\ran&=&P_\perp\frac{1}{-H_0}P_\perp V|0_N\ran\frac{\lan 0_N|\psi_N\ran}{\lan 0_N|0_N\ran}\nn\\
&&+ P_\perp\frac{1}{-H_0} P_\perp ( V-\mathcal{E}_N ) P_\perp |\psi_N\ran\, , \label{eq:PperpPsi2}
\eea
that we iterate. By inserting the resulting $P_\perp |\psi_N\ran$ into Eq.~(\ref{eq:EN2}) and by noting that $P_\perp \mathcal{E}_N |0_N\ran=0$, this gives the equation fulfilled by $\mathcal{E}_N$ as
\be
0=\lan 0_N|(V-\mathcal{E}_N ) \sum_{s=0}^\infty \left(P_\perp\frac{1}{-H_0} P_\perp (V-\mathcal{E}_N )\right)^s  |0_N\ran\, .\label{eq:mathEN}
\ee

To solve the above equation in an easy way, we split $V$ as $V_0+W$ where
\be
W=\sum_{\vq\neq \bf0}\frac{v_\vq}{2}\sum_{\vk\vk'}b_{\vk+\vq}^\dag b_{\vk'-\vq}^\dag b_{\vk'} b_\vk\, .\label{def:W}
\ee
contains all non-zero scattering processes. As $V_0$ acting on any state having $N$ bosons gives $v_{\bf0} N(N-1)/2$,  Eq.~(\ref{eq:mathEN}) ultimately appears as
\be
0=\lan 0_N|(W-\Delta_N ) \sum_{s=0}^\infty \left(P_\perp\frac{1}{-H_0} P_\perp (W-\Delta_N )\right)^s  |0_N\ran\, ,\label{eq:Delta_N}
\ee
where $\Delta_N=\mathcal{E}_N - v_{\bf0} N(N-1)/2$.

%The $P_\perp$ projection operator prevents the intermediate state to fall back onto the ground state $|\phi_N\ran$. For convenience, in the following we shall use $\lan \cdots\ran$ to denote expectation value evaluated in the $|\phi_N\ran$ state.

%In contrast to usual perturbation theory, in the Brillouin-Wigner (BW) perturbation theory the perturbation terms depend on the ground state energy $\mathcal{E}_N$ in a non-trivial way. It acts to cancel the disconnected diagrams since system energy contribution comes from connected diagrams, but it also act to cancel a few other terms that are not at all obvious.

\section{Potential expansion}

We solve Eq.~(\ref{eq:Delta_N}) as a $W$ expansion with $\Delta_N$ written as
\be
\Delta_N=\sum_{s=0}^\infty \Delta^{(s)}_N\, ,
\ee
where $s$ refers to the perturbative order in the $W$ interaction. So, $\Delta^{(s)}_N=\mathcal{E}^{(s)}_N$, except for $s=1$. \

Equation (\ref{eq:Delta_N}) readily gives $0=\Delta^{(0)}_N=\mathcal{E}^{(0)}_N$ and 0=$\Delta^{(1)}_N$; so,
\be
\mathcal{E}^{(1)}_N=  \frac{N(N-1)}{2}v_{\bf0}=\frac{\lan 0_N|V_0|0_N\ran}{\lan 0_N|0_N\ran}\, \label{eq:mathEN(1)}.
\ee
 Because $W$ operators only enter Eq.~(\ref{eq:Delta_N}), higher-order terms in interaction come from $\vq\neq\bf0$ processes.

As $\Delta_N$ is second order at least in $W$, 
%we get from Eq.~(\ref{eq:Delta_N}) 
the two next-order terms read as
\be
\Delta^{(2)}_N \lan 0_N|0_N\ran= \lan 0_N|W P_\perp\frac{1}{-H_0} P_\perp W  |0_N\ran\, ,\label{eq:DeltaN2odr}
\ee
\be
\Delta^{(3)}_N \lan 0_N|0_N\ran= \lan 0_N|W P_\perp\frac{1}{-H_0} P_\perp W P_\perp\frac{1}{-H_0} P_\perp W |0_N\ran\, ,\label{eq:DeltaN3odr}
\ee
while, for $s\ge 4$, they have a more complicated form:
\be
\Delta^{(s)}_N \lan 0_N|0_N\ran= \lan 0_N|W P_\perp\frac{1}{-H_0} P_\perp J_N^{(s)} P_\perp\frac{1}{-H_0} P_\perp W |0_N\ran\, ,
\ee
the first $J_N^{(s)}$ operators being given by
\bea
J_N^{(4)}&=& W P_\perp\frac{1}{-H_0} P_\perp W-\Delta^{(2)}_N\, ,\label{eq:JN4}\\
J_N^{(5)}&=& W P_\perp\frac{1}{-H_0} P_\perp W P_\perp\frac{1}{-H_0} P_\perp W \label{eq:JN5}\\
&&- \Delta^{(2)}_N \left( P_\perp\frac{1}{-H_0} P_\perp W+ W P_\perp\frac{1}{-H_0} P_\perp \right)- \Delta^{(3)}_N\, . \nn
\eea
From the above two equations, we see that disconnected processes exist when more than three $W$'s act inside the $|0_N\ran$ condensate, as standard for perturbative expansion. They generate overextensive contributions, which are eventually canceled out by the $\Delta_N$ parts of $J_N^{(s)}$, as shown more in details below.

\section{Some useful commutators}\label{sec:commu}
 The $\Delta^{(s)}_N$ quantities are easy to calculate with the help of the following commutator
\be
\Big[W,b^\dag_\vp\Big]_-=\sum_{\vq\neq\bf0} v_\vq b^\dag_{\vp+\vq}T_{-\vq}\, ,
\ee
where $T_{\vq}=\sum_\vk b^\dag_{\vk+\vq} b_\vk$. This excitation operator is such that
\be
\Big[T_\vq,b^\dag_\vp\Big]_-= b^\dag_{\vp+\vq}\, ,
\ee
from which we readily get
\be
T_\vq|0_N\ran=Nb^\dag_\vq |0_{N-1}\ran\, .
\ee
  By iteration, the above equations lead to
\bea
W |0_N\ran&=&\left(\Big[W,b^\dag_{\bf0}\Big]_-+b^\dag_{\bf0}W\right)|0_{N-1}\ran\nn\\
&=&\frac{N(N-1)}{2} \sum_{\vq\neq\bf0} v_\vq B^\dag_{\vq}|0_{N-2}\ran\, ,\label{eq:W0N}
\eea
where $B^\dag_{\vq}=b^\dag_\vq b^\dag_{-\vq}$ creates a finite-momentum pair from the condensate. Interaction involving such a pair follows from
\be
\Big[W,B^\dag_{\vp}\Big]_ -{=}  \sum_{\vq\neq\bf0} v_\vq B^\dag_{\vp{+}\vq}
{+}\sum_{\vq\neq\bf0} v_\vq \left(b^\dag_{\vp{-}\vq} b^\dag_{-\vp}{+}b^\dag_{\vp} b^\dag_{-\vp-\vq}\right)T_\vq\, .\label{eq:WBvp_commu}
\ee

\section{Structure of the potential expansion}

\begin{figure}[h!]
\centering
   \includegraphics
   [trim=5cm 6.8cm 5cm 20.2cm,clip,width=3in]{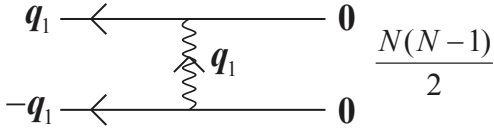}
\caption{\small One $W$ acting on $|0_N\ran$ brings one boson pair out of the condensate since $W$ contains $\vq_1\neq \bf0$ processes only. There are $N(N-1)/2$ ways to choose the two interacting bosons $\bf0$ among $N$. }
\label{fig:3}
\end{figure}

The calculation of $\Delta_N$ requires the knowledge of $s$ operators $W$ acting on the $|0_N\ran$ condensate, $\lan 0_N|W^s|0_N\ran$ containing the parts of $W^s|0_N\ran$ in which all excited bosons are back into the condensate:

%\noindent (i)
 The first operator $W$ excites two $\vq=\textbf{0}$ bosons from the condensate to become a $(\vq_1,-\vq_1)$ pair, as shown in Fig.~\ref{fig:3}. Equation (\ref{eq:W0N}) readily gives
\be
 P_\perp\frac{1}{-H_0} P_\perp W |0_N\ran= \frac{N(N-1)}{2}\sum_{\vq_1\neq\bf0}\frac{ v_{\vq_1}}{-2\va_{\vq_1}}B^\dag_{\vq_1}|0_{N-2}\ran\, .
\ee
The $N(N-1)/2$ prefactor comes from the ways to choose the two interacting bosons $\bf0$ among $N$.

%\noindent (ii) 
 A second $W$ operator acting on this excited state can do three things:
 
(i) It can induce an additional $\vq_2$  scattering between the same two bosons, leading to terms in $B^\dag_{\vq_1+\vq_2}|0_{N-2}\ran$ with the same $N(N-1)/2$ prefactor. 

(ii) It can induce an interaction with a third boson $\bf0$ from the condensate, leading to terms in $b^\dag_{-\vq_1-\vq_2}b^\dag_{\vq_1}b^\dag_{\vq_2}|0_{N-3}\ran$ with a prefactor $[N(N-1)/2](N-2)$, from the $(N-2)$ ways to choose the third boson $\bf0$.

(iii) It can excite a second boson pair $(\vq_2,-\vq_2)$ from the condensate, leading to terms in $B^\dag_{\vq_1}B^\dag_{\vq_2}|0_{N-4}\ran$ with a prefactor $[N(N-1)/2][(N-2)(N-3)/2]$.

%\noindent (iii) 
 More $W$ interactions bring more bosons $\bf0$ from the condensate into play.\
Some examples of the resulting $W^s|0_N\ran$ states are given in the supplemental material.\

As previously shown (see Eq.~(\ref{eq:mathEN(1)})), the $N$-boson energy at first order in interaction is equal to
$v_{\bf0} N(N-1)/2$, while all higher-order terms involve $\vq_i\neq\bf0$ processes through $W$ interactions. 

\begin{figure}[h!]
\centering
  \includegraphics[trim=4cm 6.5cm 4cm 18.3cm,clip,width=3in] {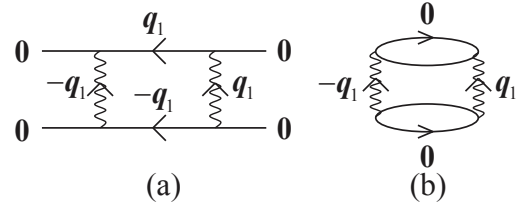}
\caption{\small Two $W$ interactions inside the $|0_N\ran$ condensate reduce to the ladder process of diagram (a). Diagram (b) is the standard bubble diagram for the same process, this Feynman diagram being obtained by closing the ``open" lines $\bf0$ of diagram (a). This process appears with a $N(N-1)/2$ prefactor due to the number of ways to choose the two interacting bosons $\bf0$ among $N$.}
\label{fig:4}
\end{figure}

$\bullet$ The second-order term, given by
\be
\mathcal{E}^{(2)}_N=\Delta_N^{(2)}=\frac{N(N-1)}{2}\sum_{\vq_1\neq\bf0}\frac{v_{-\vq_1} v_{\vq_1}}{-2\va_{\vq_1}}\ \propto\  Nn\, , \label{eq:mathEN(2)}
\ee
is visualized by the diagram of Fig.~\ref{fig:4}. 
%This quantity is extensive since its $\vq$ sum brings a volume factor when transformed into an integral.

\begin{figure}[h!]
\centering
  \includegraphics[trim=3cm 6cm 3.5cm 19.2cm,clip,width=3.2in] {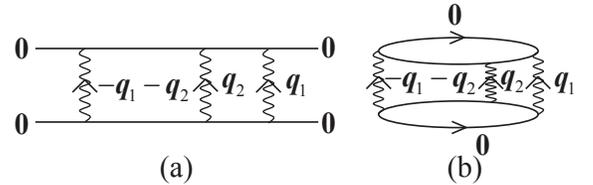}
\caption{\small Ladder process resulting from three $W$ interactions between two bosons $\bf0$ from the $|0_N\ran$ condensate. }
\label{fig:5}
\end{figure}

$\bullet$ The third-order term, $\mathcal{E}_N^{(3)}$, has two origins:

 (i) One contribution comes from the ladder process shown in Fig.~\ref{fig:5} in which $(\vq_1,\vq_2,\vq_3)$  are different from $\bf0$ but $\vq_1+\vq_2+\vq_3=\textbf{0}$. For $\vq_2$ set equal to $\vq'_2-\vq_1$, this contribution reads
\be
\frac{N(N-1)}{2} \sum_{\vq_1\not=\bf0}\sum_{\vq'_2\not=({\bf0},\vq_1)} \frac{v_{-\vq'_2}v_{\vq'_2-\vq_1}v_{\vq_1}}{(-2\va_{\vq'_2})(-2\va_{\vq_1}) }\ \propto\ Nn\, . \label{eq:termnn-1q1q'2}
\ee

(ii) The second contribution comes from the bubble processes involving three bosons $\bf0$, shown in Fig.~\ref{fig:2}. It reads
\be
2\frac{N(N-1)}{2}(N-2) \sum_{\vq_1\not=\bf0} \frac{v_{-\vq_1}v_{\vq_1}^2}{(-2\va_{\vq_1})^2 }\ \propto\ Nn^2\, ,\label{eq:termnn-1n-2q1}
\ee
the extra factor $2$ coming from the two processes in this figure which give equal contributions. 

When compared to the second-order term shown in Fig.~\ref{fig:4}, we see that a third interaction $W$ either adds one more interaction between two bubbles as in Fig.~\ref{fig:5} or adds a third bubble, as in Fig.~\ref{fig:2}. The former term, which appears with a $N(N-1)/2$ prefactor, corresponds to a renormalization of the interaction between two bosons through ladder processes. The latter term, which appears with a prefactor $N(N-1)(N-2)$, is part of the correlation energy. This term actually diverges in the small $\vq_1$ limit for $v_{\vq_1\rightarrow \bf0}\neq 0$ because it contains the $3$D integral of $(1/\vq_1^2)^2$. The $Nn^2$ dependence of this singular three-bubble process is transformed into a $Nn^{3/2}$ dependence, when joined with similar bubble terms of higher order in $W$. To understand how this happens, we must consider higher-order terms in $W$. These higher-order terms moreover bring a new feature: the appearance of overextensive contributions coming from disconnected processes. The various fourth-order terms in $W$ are given in the supplemental material, as well as the cancellation in the large $N$ limit, of their overextensive parts.\
To explicitly show the necessary cancellation of these overextensive parts through the $\Delta_N$ part of $J_N^{(s)}$ becomes more and more cumbersome when $s$ increases. From now on, we focus on extensive contributions resulting from connected diagrams.

\section{Ladder processes}

\begin{figure}[h!]
\centering
   \includegraphics[trim=3cm 3.8cm 2cm 18cm,clip,width=3.4in] {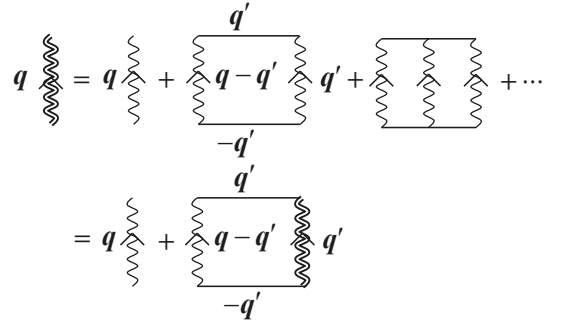}
\caption{\small  Effective scattering resulting from the repeated interaction between two bosons.}
\label{fig:13}
\end{figure}

 At any order in $W$, there are terms that come from the repeated interaction between two bosons, as shown in Figs. \ref{fig:4}, \ref{fig:5},  and \ref{fig:1}. The sum of these ladder processes leads to the effective scattering $\tilde v_\vq$ shown in Fig.~\ref{fig:13}, which is the solution of the integral equation
\be
\tilde v_\vq= v_\vq+\sum_{\vq'\not=\bf0}v_{\vq-\vq'}\frac{1}{-2\va_{\vq'}}\tilde v_{\vq'}\, .
\ee
This summation is commonly associated with the two-body scattering length $a$ defined as
\be
\lim_{\vq\rightarrow \bf0}\tilde v_\vq=4\pi\frac{a}{mL^3}\, .
\ee

It is possible to rewrite the energy term coming from any number of interaction between two bosons taken from the condensate, that is, all processes like the ones of Figs. \ref{fig:4}, \ref{fig:5},  and \ref{fig:1} plus the first-order term given in Eq.~(\ref{eq:mathEN(1)}), in terms of the scattering length $a$, as
\be
\frac{N(N-1)}{2}\tilde v_{\bf0}= N\frac{2\pi}{m}na\, ,
\ee
 in agreement with the first term of Eq.~(\ref{eq:MFenergy}).\

A similar interaction renormalization exists between bubbles. It leads to replacing $v_\vq$ with $\tilde v_\vq$. Actually, this simple procedure is only valid in the small $\vq$ limit, which is when bubble processes are most singular; for larger $\vq$'s, the interaction renormalization is affected by the surrounding excited states.

\begin{figure}[h!]
\centering
  \includegraphics[trim=3cm 2.8cm 2.7cm 17.5cm,clip,width=3.4in] {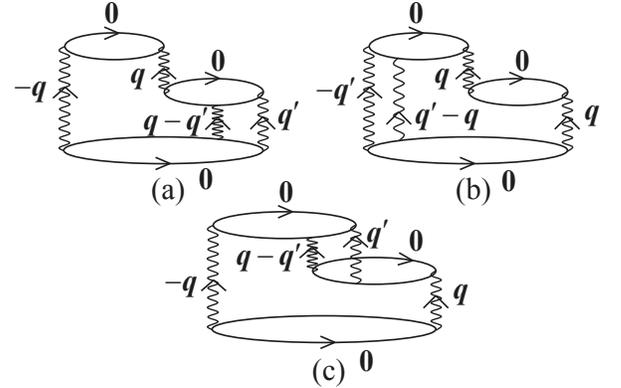}
\caption{\small  Repeated interaction between three bubbles leading to Eq.~(\ref{eq:vqrenormalized}).}
\label{fig:14}
\end{figure}

To understand this important point, let us consider three-bubble processes.
The one involving three $v_{\vq}$'s, shown in Fig.~\ref{fig:2}, leads to the contribution given by Eq.~(\ref{eq:termnn-1n-2q1}).
%\be
%2\frac{N(N-1)}{2}(N-2) \sum_{\vq_1\not=\bf0} \frac{v_{-\vq_1}v_{\vq_1}^2}{(-2\va_{\vq_1})^2 }\, ,\label{eq:termnn-1n-2q1}
%\ee
%the extra factor $2$ coming from the two processes shown in Fig.~\ref{fig:2}(a,b). We will see that the prefactor $2$ is crucial to properly sum up bubble diagrams. We will give a general way to count these equivalent but topologically different bubble diagrams.\
These three $v_\vq$ interactions are also renormalized by adding more $W$ interactions, as shown in Fig.~\ref{fig:14}. The three bubble diagram then transforms up to fourth order in $W$ into
\bea
N(N-1)(N-2)\sum_{\vq\not=\bf0}\frac{ 1}{(-2\va_\vq)^2 }\bigg(\! v_\vq+\sum_{\vq'\not=\bf0}v_{\vq-\vq'}\frac{v_{\vq'}}{-2\va_{\vq'}}\bigg)^2\nn\\
\times \bigg(\! v_\vq+\sum_{\vq'}v_{\vq-\vq'}\frac{v_{\vq'}}{-\va_\vq-\va_{\vq'}-\va_{\vq-\vq'}}\bigg)\nn\\
\sim N(N-1)(N-2)\sum_{\vq\not=\bf0}\frac{ \tilde v_\vq^2 \tilde v'_\vq}{(-2\va_\vq)^2 }\hspace{2cm}\label{eq:vqrenormalized}
\eea
The $\tilde v'_{\vq}$ interaction associated with Fig.~\ref{fig:14}(c) does not read exactly the same as the $\tilde v_\vq$ interactions coming from the diagrams in Fig.~\ref{fig:14}(a,b). However, the dominant contribution of these three-bubble processes comes from small $\vq$'s because of the divergence of the $1/(-2\va_\vq)^2$ factor it contains; for such momenta, $\tilde v'_\vq\simeq \tilde v_\vq$. \

%Similar interaction renormalization exists for higher-order bubble processes. They lead to replacing $v_\vq$ with $\tilde v_\vq$. We wish to stress that this simple procedure is valid in the small $\vq$ limit; so, it is appropriate to derive the most singular terms of the bubble processes. This is not so for large $\vq$'s, because the interaction renormalization is then affected by the surrounding excited states, as clearly seen from Eq.~(\ref{eq:vqrenormalized}).

\section{Origin of the $n^{3/2}$ singularity}

We now consider the $N$-boson correlation energy, \emph{i.e.}, the terms coming from interactions between more than two bosons, and we explain why the dominant correlation energy term does not depend on density as $n^2$, but as $n^{3/2}$. This non-perturbative result originates from small-$\vq$ divergences such as those in Eqs.~(\ref{eq9}) and (\ref{eq:termnn-1n-2q1}). The standard procedure to remove these singularities is to sum up similar divergent terms.  \

In order to explicitly show how this summation removes the singularities, let us consider the lowest-order terms represented by the diagrams of Figs.~\ref{fig:4}, \ref{fig:2}, and 21 of the supplemental material. In this set of processes, the system has one pair $(\vq,-\vq)$ excited from the $\vk={\bf 0}$ condensate, along all intermediate steps, this pair being ultimately de-excited back into the condensate.
These diagrams lead to
\begin{subeqnarray}
\label{eq:energyDelta}
\lefteqn{\Delta'_N=}\nn\\
&& \frac{N(N-1)}{2} \sum_{\vq\neq {\bf 0}}\frac{v_\vq^2}{-2\va_\vq}+2\frac{N(N-1)}{2}(N-2)\sum_{\vq\neq {\bf 0}}\frac{v_\vq^3}{(-2\va_\vq)^2}\nn\\
& &+4\frac{N(N-1)}{2}(N-2)(N-3)\sum_{\vq\neq {\bf 0}}\frac{v_\vq^4}{(-2\va_\vq)^3}+\cdots\slabel{eq:energyDeltaa}\\
&&\simeq\frac{1}{2}\sum_{\vq\neq {\bf 0}}\frac{(Nv_\vq)^2}{-2\va_\vq-2Nv_\vq}\, .\slabel{eq:energyDeltab}
\end{subeqnarray}
for $N$ large. The effect of summing up this series is clear: it brings an interaction term to the pair kinetic energy, which stays finite in the small $\vq$ limit, thus serving as a small-$\vq$ energy cut-off, provided that $v_{\bf0}$ is finite.

Of course, the above result follows from appropriate prefactors. Let us show that bubble processes involving $s\geqslant3$ interactions lead to
\be
2^{s-2}\frac{N(N-1)}{2}(N-2)\cdots (N-s+1)\sum_{\vq\neq {\bf 0}}\frac{v_\vq^s}{(-2\va_\vq)^{s-1}}\,\label{eq:energy-renorm} .
\ee
Since this term involves $s$ bosons, its $N$ factor is easy to understand: $N(N-1)/2$ comes from the number of ways to choose the first pair of bosons $\bf0$ from $N$ identical bosons $\bf0$ in the condensate, as shown in Fig.~\ref{fig:3}. The other $N$ factors come from adding additional bosons $\bf0$ to this initial pair, the number of bosons that remain in the condensate decreasing as $(N-2)$, $(N-3)$, $\ldots$\

\begin{figure}[h!]
\centering
   \includegraphics[trim=3cm 6.5cm 3cm 19cm,clip,width=3.4in] {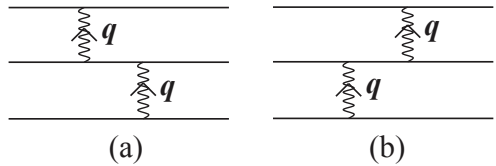}
\caption{\small The $\mathcal{U}$ operator adds an up-line to the diagram of Fig.~\ref{fig:3}, as in (a), while the $\mathcal{D}$ operator adds a down-line, as in (b). }
\label{fig:15}
\end{figure}
The $2^{s-2}$ prefactor is more tricky. It comes from topologically different diagrams, such as the two diagrams in Fig.~\ref{fig:2} that involve $s=3$ interactions. Let us provide a systematic way to generate this $2^{s-2}$ prefactor. We start with the $\vq$ interaction between two boson lines, shown in Fig.~\ref{fig:3}, and we introduce two operators $\mathcal{U}$ and $\mathcal{D}$ : the $\mathcal{U}$ operator adds an up-line connected to the upper boson line by a $\vq$ interaction, as in Fig.~\ref{fig:15}(a); the $\mathcal{D}$ operator adds a down-line connected to the lower boson line by a $\vq$ interaction, as in Fig.~\ref{fig:15}(b). To construct the diagrams having $s$ bubbles, we apply $(\mathcal{U}+\mathcal{D})^{s-2}$ to the two-boson diagram of Fig.~\ref{fig:3} and we close the uppermost and lowest boson lines by a $-\vq$ interaction. This binomial expansion leads to a prefactor $2^{s-2}$, since each term in the expansion corresponds to topologically different process, while the integral is the same. The corresponding third-order and fourth-order bubble diagrams are respectively shown in Fig.~\ref{fig:2}, and 21.

By rescaling $\va_\vq$ in Eq.~(\ref{eq:energyDeltab}) as $2Nv_{\bf 0}x$, we get
\be
\Delta'_N\simeq \frac{1}{2}\left(\frac{L}{2\pi}\right)^3 4\pi m\sqrt{2m}(2Nv_{\bf 0})^{5/2} S_N\label{eq:Delta'_N}\, ,
\ee
where, for $v_\vq$ taken equal to $v_{\bf 0}$ when $0\leq\va_\vq\leq\Omega$ and $0$ otherwise, in order to insure the large-$\vq$ convergence of the $\vq$ sums which is normally insured by the natural decrease of $v_{\bf q}$, the $S_N$ sum appears as
\bea
S_N&=& -\frac{1}{4}\int_0^{\Omega/2Nv_{\bf 0}} dx\,\frac{\sqrt{x}}{2x+1}\nn\\
&=& -\frac{1}{4}\left[\sqrt{\frac{\Omega}{2Nv_{\bf 0}}}-\frac{1}{\sqrt{2}}\tan^{-1}\sqrt{\frac{\Omega}{Nv_{\bf 0}}}\right]\, .\label{eq:SN}
\eea
The first term of $S_N$, which produces a $v_{\bf 0}^2$ term in $\Delta'_N$, comes from the first term in Eq.~(\ref{eq:energyDeltaa}). It should be subtracted off because it is already included in the ladder series that produce the part of $\Delta_N$ linear in density. By letting the potential cut-off $\Omega$ go to infinity in the second term of $S_N$, we get a  contribution in $v_{\bf 0}^{5/2}$ to $\Delta'_N$, that can be written in terms of the scattering length at first order in interaction, $a_1\simeq mL^3 v_{\bf0}/4\pi$, as
\be
N \frac{4\pi\sqrt{2\pi}}{ma_1^2}  (na_1^3)^{3/2}\, .\label{DeltaNreslt}
\ee

Difference of this result from the second term of the $N$-boson energy in Eq.~(\ref{eq:MFenergy}) is two-fold: (i) it reads in terms of the first-order scattering length $a_1$, instead of the full scattering length $a$; (ii) its numerical factor is different.

\noindent (i) To get the full scattering length $a$, instead of $a_1$, in the above equation is quite easy: we just have to note that ladder processes also exist between bubbles, as shown in Fig.~\ref{fig:14}. In the small $\vq$ limit where bubble contributions are singular, they lead to replacing $v_{\bf0}$ in Eqs.~(\ref{eq:Delta'_N}) and (\ref{eq:SN}) by $\tilde v_{\bf0}$, that is replacing $a_1$ by $a$ in Eq.~(\ref{DeltaNreslt}).\

\noindent (ii) The numerical factor in Eq.~(\ref{DeltaNreslt}) differs from the one of the $a^{5/2}$ term in Eq.~(\ref{eq:MFenergy}), by $4\%$ only\cite{BruecknerPR1957}. This evidences that the dominant contribution to the $N$-boson correlation energy comes from the repeated excitation of just one boson pair from the condensate. The missing small contribution comes from processes in which more than one boson pair $(\vq,-\vq)$ are excited from the condensate. Actually, the correlation energy appears as
\begin{subeqnarray}
\label{eq:Corr_energy}
\Delta^{corr}_N&\simeq& 1\frac{1}{2}\sum_{\vq\neq {\bf 0}}\frac{(Nv_\vq)^2}{-2\va_\vq-2Nv_\vq}+1\frac{1}{2}\sum_{\vq\neq {\bf 0}}\frac{(Nv_\vq)^4}{(-2\va_\vq-2Nv_\vq)^3}\nn\\
&&+2\frac{1}{2}\sum_{\vq\neq {\bf 0}}\frac{(Nv_\vq)^6}{(-2\va_\vq-2Nv_\vq)^5}+\cdots\slabel{Corr_energya}\\
&=& \frac{1}{2}\sum_{\vq\neq {\bf 0}}\sum_{s=1}^\infty f_{s}\frac{(Nv_\vq)^{2s}}{(-2\va_\vq-2Nv_\vq)^{2s-1}} \slabel{Corr_energyb}
\end{subeqnarray}

$\bullet$ The first term of Eq.~(\ref{Corr_energya}) comes from the excitation of one boson pair, as given by Eq.~(\ref{eq:energyDeltab}).\

\begin{figure}[h!]
\centering
   \includegraphics[trim=3cm 5.5cm 3cm 18cm,clip,width=3.4in] {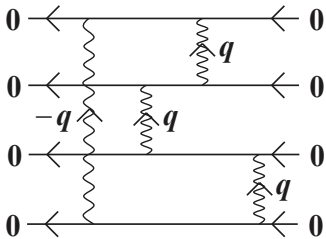}
\caption{\small Process involving the excitation of two boson pairs $(\vq,-\vq)$.  }
\label{fig:16}
\end{figure}

 $\bullet$ The second term comes from the process shown in Fig.~\ref{fig:16}, in which two boson pairs are excited and then de-excited, the energy denominator $-2\va_\vq$ being replaced by $-2\va_\vq-2Nv_\vq$, to account for the renormalization from bubble processes, as shown in Eq.~(\ref{eq:energyDeltaa}). Since at least one excited boson pair must remain in the intermediate steps, there is only one way to destroy the two pairs: they are created first and then destroyed.\

\begin{figure}[h!]
\centering
   \includegraphics[trim=1.5cm 3.6cm 1cm 17.5cm,clip,width=3.5in] {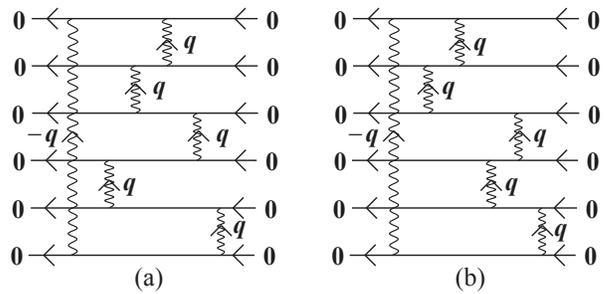}
\caption{\small  Processes involving the excitation of three boson pairs $(\vq,-\vq)$. Process (a) excites three pairs and then destroy them all. Process (b) first excites two pairs, then destroys one pair, then excites another pair, and finally destroys them all. }
\label{fig:17}
\end{figure}

$ \bullet $ The third term of Eq.~(\ref{Corr_energya}) comes from the two processes shown in Fig.~\ref{fig:17}, in which three boson pairs are excited and then de-excited. These two processes correspond to the two ways to destroy three boson pairs: (i) one can excite three bosons pairs and then destroy them all, as in Fig.~\ref{fig:17}(a); (ii) one can excite two boson pairs, destroy a pair, excite another pair, and finally destroy them all, as in Fig.~\ref{fig:17}(b). 

Following this simple rule, it is easy to count all possible ways of destroying $s$ excited pairs. The $f_{s}$ prefactor in Eq.~(\ref{Corr_energyb}) represents the number of ways that $s$ excited boson pairs can be destroyed. \
Brueckner and Sawada\cite{BruecknerPR1957} have shown that this $f_{s}$ factor is just the numerical factor of the generating function
\be
\frac{1}{2}\left\{ 1-\sqrt{1-4x}\right\}=\sum_{s=1}^\infty f_s x^s
\ee
Using this function, it is possible to show that the energy given in Eq.~(\ref{eq:Corr_energy}) is nothing but the correlation energy obtained from the mean-field approach, in this way recovering the precise numerical prefactor for the $n^{3/2}$ term given in Eq.~(\ref{eq:MFenergy}).

\section{State-of-the-art for composite bosons}

The major problem when dealing with composite bosons (``cobosons") comes from the fact that, due to possible fermion exchanges, there is no way to assign a given fermion pair to a composite boson. As a direct consequence, there is no way to write an effective Hamiltonian with an interaction potential between cobosons that is valid beyond first-order processes, in spite of what all bosonization procedures intend to do. So, previous approaches used for elementary bosons, which are based on the existence of a boson-boson potential such as the one of Eq.~(\ref{def:V1}), cannot be used for composite boson systems. Of course, it is always possible to stay with the potential between elementary fermions, as done in refs.~\onlinecite{Leyronas2007,Alzetto2013}. However, the ladder processes that lead to the formation of a composite boson, have to be selected and summed up first, before considering  interaction between composite objects. In this regard, the coboson many-body formalism\cite{moniqPhysRep} proposed a decade ago is far simpler because it avoids the handling of ladder processes associated with composite boson formation. \

\begin{figure}[h!]
\begin{center}
   \includegraphics[trim=2.5cm 4.2cm 2cm 17.4cm,clip,width=3.4in] {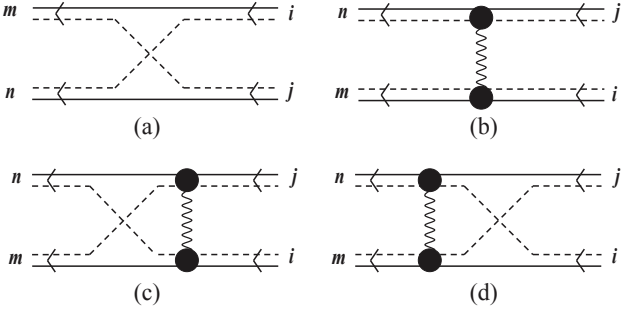}
   \caption{\small (a) The Pauli scattering $\lambda(_{mi}^{\, nj})$. (b) The direct Coulomb scattering $\xi(_{mi}^{\, nj})$. (c) The ``in" exchange Coulomb scattering $\xi^{in}(_{mi}^{\, nj})$. (d) The ``out" exchange Coulomb scattering $\xi^{out}(_{mi}^{\, nj})$. }
   \label{fig:18}
\end{center}
\end{figure}

The dominant way\cite{moniqPhysRep} two cobosons interact is through the Pauli scattering induced by the Pauli exclusion principle, as represented by the Shiva diagram in Fig.~\ref{fig:18}(a). Since, in it, no interaction takes place, this dimensionless scattering is by construction missed by all bosonized Hamiltonians, because in them, only enter energy-like quantities.\

Attempts have been made to reach the $N$-coboson energy linear in density in the case of cold atom dimers within a contact potential\cite{Petrov2004,Alzetto2013}, and in the case of semiconductor excitons\cite{Keldysh1968,Shiau2015} within the long-range Coulomb potential. The cold atom problem is rather simple because in the linear term in density, only enter two bosonic atoms; so, this linear term can be obtained by numerically solving a four-fermion problem. The next-order density term, that is, the correlation energy, is far more tricky to get because it requires solving a full many-body problem. Indeed, it is reasonable to expect that, as in the case of elementary bosons, this term has a singular density dependence that comes from more than three fermion pairs. So, a numerical brute-force calculation is hopeless. In the case of cold atomic gases with a contact potential, it has been shown\cite{Leyronas2007}, using a procedure similar to the one proposed by Lee-Huang-Yang, that the $N$-dimer energy is also given by Eq.~(\ref{eq:MFenergy}), where $a$ is the scattering length between two dimers.

These procedures, hard to extend to the long-range Coulomb potential, fail to address the main physical problem when dealing with composite bosons: how do fermion exchanges enter the result? The repeated interaction between \textit{two} cobosons must contain the direct Coulomb scattering shown in Fig.~\ref{fig:18}(b), as well as the two exchange Coulomb scatterings shown in Fig.~\ref{fig:18}(c,d). It can also contain the Pauli scattering induced by fermion exchange, multiplied by the difference of the coboson energies of the process at hand in order to get an energy-like quantity. So, we expect the effective scattering appearing in the ladder processes between two cobosons to read\cite{Shiau2015}
\bea
\zeta\left(\begin{smallmatrix}
m& i \\ n& j \end{smallmatrix}\right)&=&(\cdots)\xi\left(\begin{smallmatrix}
m& i \\ n& j \end{smallmatrix}\right)+(\cdots)\xi^{in}\left(\begin{smallmatrix}
m& i \\ n& j \end{smallmatrix}\right)+(\cdots)\xi^{out}\left(\begin{smallmatrix}
m& i \\ n& j \end{smallmatrix}\right)\nn\\
&&+(\cdots)(E_m+E_n-E_i-E_j)\lambda\left(\begin{smallmatrix}
m& i \\ n& j \end{smallmatrix}\right)
\eea

It is far from obvious that the set of exchange processes between two cobosons will appear just the same in the case of many cobosons. One simple reason is that Pauli scatterings between three cobosons $\lambda\left(\begin{smallmatrix}
m& i \\ n& j \\ l& k\end{smallmatrix}\right)$  or more also exist\cite{moniqPhysRep}. So, even if the next-order term in density still behaves as $n^{3/2}$, its prefactor may not solely depend on the scattering length between just two cobosons. The understanding, through the present work, that the $n^{3/2}$ singular dependence comes from the accumulation of the same non-zero momentum transfer excitations from the condensate provides a great clue and a guidance for suggesting the series of Shiva diagrams that should lead to the most singular term in the $N$-composite boson correlation energy, and for possibly confirming the result obtained in the case of atomic dimers\cite{Leyronas2007}.\

\section{Conclusion}

We propose a compact perturbative procedure to derive the energy of $N$ interacting bosons obtained by Brueckner and Sawada\cite{BruecknerPR1957} and by Lee, Huang, and Yang\cite{LY1957,LHY1957}, through the summation of an infinite series of diagrams of the ``ladder" type to get the scattering length and of the ``bubble" type to get the correlation energy --- which depends singularly on the density $n$ as  $n^{3/2}$ while its density dependence is (ln $n$) for fermions.
One advantage of the present procedure is that it provides a unified way to derive the singular correlation energy of fermions and bosons in terms of $\vq\neq \bf0$ excitations from their corresponding non-interacting ground states: the Bose-Einstein condensate for bosons and the Fermi sea for fermions. The other advantage is that it gives a physical picture for how the singular behavior arises from the accumulation of $\vq\neq \bf0$ excitations. This understanding will serve as a valuable guide to study composite  boson systems such as Bose-Einstein condensates of cold atoms or semiconductor excitons, by selecting appropriate Shiva diagrams having similarity with ``bubble" Feynman diagrams.

\section*{Acknowledgments}
This work is supported by the Headquarters of University Advancement at National Cheng-Kung university,  National Science Council of Taiwan under Contract No. NSC 101-2112-M-001-024-MY3, and Academia Sinica, Taiwan. M.C. wishes to thank the National Cheng Kung University and the National Center for Theoretical Sciences (South) for invitations. S.-Y. S. also wishes to thank the Institut des NanoSciences de Paris for invitation.

\section{Supplemental material}

\renewcommand{\thesection}{\mbox{Appendix~\Roman{section}}} %\section{Appendix}
\setcounter{section}{0}

\renewcommand{\theequation}{\mbox{A.\arabic{equation}}} %\section{Appendix}
\setcounter{equation}{0} %

\mbox{}\\
{\bf Appendix A. Brueckner-Sawada approach}

\mbox{}

Brueckner and Sawada proposed a mean-field approach to the energy of $N$ interacting elementary bosons. Their boson Hamiltonian reads $H=H_0+V$ with $H_0$ given in Eq.~(3) and $V$ given in Eq.~(4), the operator $c_\vk^\dag$ being the elementary boson creation operator $b^\dag_\vk$. They performed a mean-field treatment in which the creation and destruction operators $b^\dag_{\bf0}$ and $b_{\bf0}$ are replaced by a scalar $\sqrt{N_{\bf0}}$, with $N_{\bf0}$ being the number of bosons $\bf0$ in the condensate. The total number of bosons then reads
\be
N=N_{\bf0}+\sum_{\vk\neq\bf0}N_\vk=N_{\bf0}+\sum_{\vk\neq\bf0} b^\dag_\vk b_\vk\, .\label{app:NN0Nk}
\ee

In the $N\simeq N_{\bf0}$ regime, the $V$ dominant terms have four and two boson $\bf0$ operators, but not three due to momentum conservation. The part in $v_{\bf0}$ reads as
\bea
&&\frac{v_{\bf0}}{2}\bigg\{ b^\dag_{\bf0} b^\dag_{\bf0} b_{\bf0}b_{\bf0}+ b^\dag_{\bf0}\Big( \sum_{\vk\neq\bf0} b^\dag_\vk b_\vk +\sum_{\vk'\neq\bf0} b^\dag_{\vk'} b_{\vk'} \Big) b_{\bf0}\bigg\}\nn\\
&&\simeq\frac{v_{\bf0}}{2} N^2\, ,
\eea
due to Eq.~(\ref{app:NN0Nk}), while the part in $v_{\vq\neq \bf0}$ is given by
\bea
&&\bigg\{ \sum_{\vq\neq\bf0} \frac{v_{\vq}}{2}b^\dag_\vq b^\dag_{-\vq} \bigg\} b_{\bf0} b_{\bf0}+ b^\dag_{\bf0} b^\dag_{\bf0}\bigg\{ \sum_{\vq\neq\bf0} \frac{v_{\vq}}{2} b_\vq b_{-\vq} \bigg\} \nn\\
&&+b^\dag_{\bf0} \bigg\{ \sum_{\vq\neq\bf0} \frac{v_{\vq}}{2}\Big(b^\dag_\vq b_{\vq}+b^\dag_{-\vq} b_{-\vq} \Big) \bigg\} b_{\bf0}\, .
\eea

Since $\va_{\vk=\bf0}=0$, we end with a mean-field Hamiltonian given by
\be
H^{(MF)}=\frac{v_{\bf0}}{2}N^2+\frac{1}{2}\sum_{\vq\neq\bf0} h_\vq\, ,
\ee
 where $h_\vq$ appears as
\be
h_\vq=\tilde \va_{\vq}\Big( b^\dag_\vq b_\vq +b^\dag_{-\vq} b_{-\vq} \Big) +\nu_\vq \Big(b^\dag_\vq b^\dag_{-\vq}+b_{\vq} b_{-\vq} \Big)\, .
\ee
The effect of the $\vk=\bf0$ condensate is to renormalize the $\vq\neq \bf0$ interaction as $\nu_\vq=Nv_\vq$ and to dress the $\vq\neq\bf0$ boson energy as $\tilde \va_\vq=\va_\vq+\nu_\vq$, the interaction ultimately appearing as creation or destruction of $(\vq,-\vq)$ boson pairs.\

The $h_\vq$ Hamiltonian is easy to diagonalize in terms of Bogoliubov-like operators
\be
\beta_\vq^\dag= b^\dag_\vq \cosh \theta_\vq +b_{-\vq} \sinh \theta_\vq\, .
\ee
For $\tanh 2\theta_\vq= \nu_\vq/\tilde \va_\vq$, we find that $h_\vq$ can be written as
\be
h_\vq =-\tilde \va_\vq\left(1-\frac{1}{\cosh 2\theta_\vq }\right)+\frac{\tilde \va_\vq}{\cosh 2\theta_\vq}\Big(\beta_\vq^\dag\beta_\vq+\beta_{-\vq}^\dag\beta_{-\vq}\Big)\, .
\ee
So, the mean-field Hamiltonian ultimately reads
\be
H^{(MF)}= E_N^{(MF)}+\sum_{\vq\neq\bf0}\sqrt{\tilde \va_\vq^2-\nu_\vq^2}\, \beta_\vq^\dag\beta_\vq\, ,
\ee
where $\beta_\vq^\dag$ creates excitation with energy $\sqrt{\tilde \va_\vq^2-\nu_\vq^2}$ from the mean-field ground state. The mean-field ground state energy is given by
\be
E_N^{(MF)}=\frac{v_{\bf0}}{2}N^2+\frac{1}{2}\sum_{\vq\neq\bf0} \Big( \sqrt{\tilde \va_\vq^2-\nu_\vq^2}-\tilde \va_\vq \Big)\, .\label{app:meanfieldEN}
\ee

The simplest way to estimate this energy is to force the scattering length into the problem. It reads, up to second order in interaction,
\be
4\pi\frac{a}{mL^3}=\tilde v_{\bf0}\simeq v_{\bf0}+\sum_{\vq\neq\bf0} v_{-\vq}\frac{1}{-2\va_\vq}v_\vq\, .
\ee
This allows us to write the ground state energy as
\be
E_N^{(MF)}=N\frac{2\pi}{m}na+E'_N\, .
\ee
with $n=N/L^3$ and $E'_N$ given by
\bea
E'_N&=&\frac{1}{2}\sum_{\vq\neq\bf0} \Big( \sqrt{\tilde \va_\vq^2-\nu_\vq^2}-\tilde \va_\vq +\frac{\nu_\vq^2}{2\va_\vq}\Big)\label{app:EprimeN}\\
&=&\frac{1}{2}\left(\frac{L}{2\pi}\right)^3 4\pi m\sqrt{2m}(2Nv_{\bf0})^{5/2}K\nn \, .
\eea
For $v_\vq$ taken equal to $v_{\bf0}$ up to $\vq$ infinite and for $\va_\vq$ rescaled as $2N v_{\bf0}x$, the $K$ integral reads as
\be
K=\int_0^\infty dx \sqrt{x}\bigg(\sqrt{x^2+x}-(x+\frac{1}{2})+\frac{1}{8x}\bigg)\, .
\ee
By noting that the primitive of this integral is given by
\be
K(x) =\frac{2}{5}\Big\{(x+1)^{5/2}-x^{5/2}\Big\}-\frac{2}{3}\Big\{(x+1)^{3/2}+\frac{1}{2}x^{3/2}\Big\}+\frac{\sqrt{x}}{4}\, ,
\ee
so that $K(0)=-4/15$ and $K(\infty)=0$, we ultimately find the $N$-boson energy in the mean-field approximation as
\be
\frac{E_N^{(MF)}}{N}=\frac{2\pi}{m}n\left(a+\frac{128}{15\sqrt{\pi}}a_1 \sqrt{na_1^3}\right)\, .
\ee
This result differs from Eq.~(1) through the fact that the $n^{3/2}$ term depends on the scattering length $a_1$ associated with the bare potential $v_{\bf0}$, instead of $a$ associated with the interaction scattering dressed by ladder processes. Approaches beyond the mean-field approximation are required to transform $a_1$ into the full scattering length $a$ in the correlation term.

\renewcommand{\theequation}{\mbox{B.\arabic{equation}}} %\section{Appendix}
\setcounter{equation}{0} %

\mbox{}\\
{\bf Appendix B. Lee-Huang-Yang approach}

\mbox{}

The Lee-Huang-Yang approach\cite{LY1957,LHY1957} takes the force between two bosons as a hard-sphere interaction. This hard-sphere interaction is further approximated by a pseudo-potential\cite{HY1957}
\be
V^{(pseudo)}=\frac{4\pi a}{m}\sum_{i>j}\delta(\vr_i-\vr_j)\frac{\partial}{\partial r_{ij}}r_{ij}\, ,\label{app:pseudo-pote}
\ee
where $\vr_i$ is the position of the $i$ boson and $r_{ij}$ is the distance between $i$ and $j$ bosons. For such a hard-sphere interaction between two bosons with small momenta $\vk$ ($|\vk|a\ll 1$), the parameter $a$ physically corresponds to the hard sphere diameter. When the potential in Eq.~(\ref{app:pseudo-pote}) is used in the case of just two bosons, one finds that $a$ corresponds to the {\it full} two-body scattering length. So,  the great advantage of the above pseudo-potential is to force the full scattering length into the problem from the very beginning. Besides the delta function $\delta(\vr_i-\vr_j)$ describing a contact force, the peculiar form of $(\partial/\partial r_{ij})r_{ij}$ reflects the boundary condition of hard spheres at $r=a$, below which the wave function must reduce to zero; it guarantees that there is no $1/r$ singularity when $r\rightarrow 0$. When written in second quantization, the potential in Eq.~(\ref{app:pseudo-pote}) appears as
\be
V^{(pseudo)}=\frac{2\pi a}{m L^3}\lim_{r\rightarrow 0}\frac{\partial}{\partial r}r \sum_{\vk' \vk} \sum_{\vq} e^{i\vq\cdot \vr} b^\dag_{\vk+\vq} b^\dag_{\vk'-\vq} b_{\vk'} b_{\vk}\, .
\ee

The Hamiltonian of hard-sphere bosons is then treated in a mean-field approximation like in the Brueckner-Sawada approach. By singling out the dominant interactions involving four $\bf0$ bosons and two $\bf0$ bosons, the approximated Hamiltonian reduces to
\bea
H&=&\frac{2\pi aN^2}{m L^3}+\frac{1}{2} \lim_{r\rightarrow 0} \frac{\partial}{\partial r} r \sum_{\vq\neq \bf0}e^{i\vq\cdot \vr}\nn\\
&&\times \left\{\left(\va_\vq+\frac{4\pi aN}{m L^3}\right) \big(b_\vq^\dag b_\vq +b_{-\vq}^\dag b_{-\vq} \big)\right.\nn\\
&&\left.+\frac{4\pi aN}{m L^3} \big(b_\vq^\dag b^\dag_{-\vq} +b_{\vq} b_{-\vq} \big) \right\}\, ,
\eea
which can be exactly diagonalized by using a standard Bogoliubov procedure. We then find that the ground-state energy reads as Eq.~(\ref{app:EprimeN}), with $\tilde \va_\vq$ replaced by $\va_\vq+4\pi aN/m L^3$, and $\nu_\vq$ replaced by $4\pi aN/m L^3$. In addition to having the full scattering length $a$ appearing into the problem at no cost, another advantage with the pseudo-potential (\ref{app:pseudo-pote}) is that the last term of Eq.~(\ref{app:EprimeN}), which now reads $(4\pi aN/m L^3)^2/2\va_\vq$, appears in a natural way to remove the $1/q^2$ singularity that is hidden in the first term.

\renewcommand{\theequation}{\mbox{C.\arabic{equation}}} %\section{Appendix}
\setcounter{equation}{0} %

\mbox{}\\
{\bf Appendix C. $W^s$ acting on $|0_N\ran$}

\mbox{}

%\section{$W^n$ acting on $|0_N\ran$\label{app:sec3}}

$\bullet$ A first $W$ potential acting on $|0_N\ran$ excites one boson pair from the condensate (see Eq.~(39))
\be
  W |0_N\ran= \frac{N(N-1)}{2}\sum_{\vq_1\neq\bf0}v_{\vq_1}B^\dag_{\vq_1}|0_{N-2}\ran\, ,
\ee
 where $ B^\dag_{\vq_1}=b^\dag_{\vq_1}b^\dag_{-\vq_1}$ creates a boson pair.
 The resulting state is shown in Fig.~4.

\begin{figure}[h!]
\centering
   \includegraphics[trim=1cm 2.5cm 1cm 15cm,clip,width=3.5in] {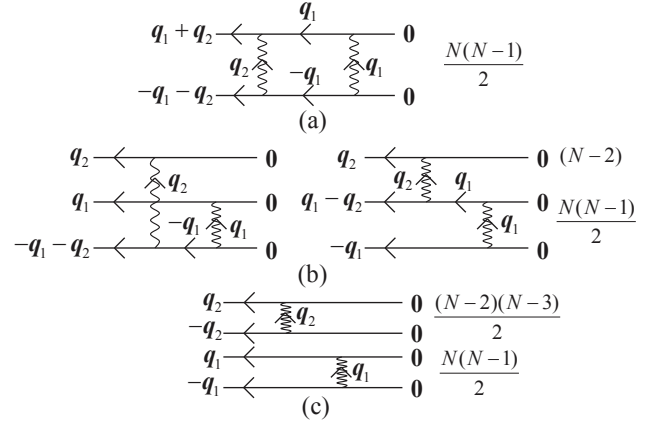}
\caption{\small A second $W$ acting on the state of Fig.~4 can act on the excited pair, as in (a), or can involve one or two additional bosons $\bf0$ from the condensate, as in (b) and (c). }
\label{fig:c1}
\end{figure}

$ \bullet$ A second $W$ acting on this one-boson-pair excited state gives, using Eqs.~(39) and (40),
\bea
\lefteqn{W B^\dag_{\vq_1}|0_{N-2}\ran =\sum_{\vq_2\neq\bf0}v_{\vq_2}B^\dag_{\vq_1+\vq_2}|0_{N-2}\ran}\\
&&{+}(N-2)\sum_{\vq_2\neq\bf0}v_{\vq_2}\left(b^\dag_{\vq_1}b^\dag_{-\vq_1-\vq_2 }{+}b^\dag_{\vq_1-\vq_2}b^\dag_{-\vq_1} \right)b^\dag_{\vq_2} |0_{N-3}\ran\nn\\
&&{+}\frac{(n-2)(N-3)}{2}  B^\dag_{\vq_1}\sum_{\vq_2\neq\bf0}v_{\vq_2} B^\dag_{\vq_2} |0_{N-4}\ran\nn\, .
\eea
This state is shown in Fig.~\ref{fig:c1}.
It contains states having two, three and four bosons excited from the condensate, with prefactors $N(N-1)/2$, $[N(N-1)/2](N-2)$, and $[N(N-1)/2][(N-2)(N-3)/2]$, respectively, which come from the number of ways to choose these bosons $\bf0$ among $N$.\

 %\subsection{Third-order term}

 %The energy at third order in $W$ reduces to Eq.~(\ref{eq:DeltaN3odr}). It is obtained by adding one $\vq_3\neq\bf0$ interaction to the second-order processes shown in Fig.~\ref{fig:3} and by projecting the resulting state on $|0_N\ran$. Instead of performing a brute-force calculation of this third-order term using the commutators given in Sec.~\ref{sec:commu}, let us derive it through diagrams. This not only helps to understand the physics at hand but also paves the way to easily grasp the structure of higher-order terms.

 $\bullet$ We now consider a third $W$ acting on the states of Fig.~\ref{fig:c1} having two, three, and four excited bosons, and we successively consider what $W$ does on these states.
\begin{figure}[h!]
\centering
   \includegraphics[trim=3.5cm 2cm 3.2cm 15cm,clip,width=2.9in] {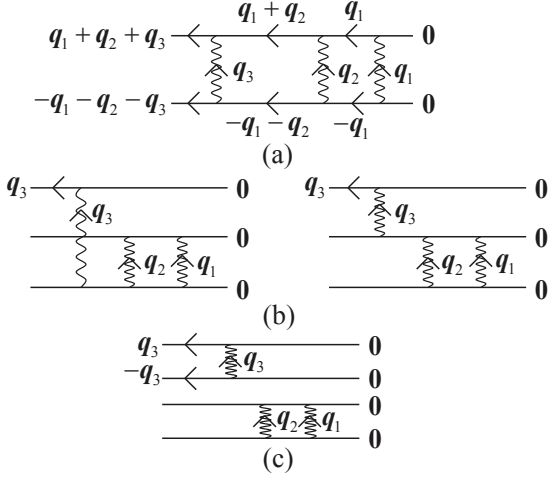}
\caption{\small A third $\vq_3\neq\bf0$ added to the processes in Fig.~\ref{fig:c1} can involve the same two bosons $\bf0$ as in (a), a third boson $\bf0$ as in (b), or excite another boson pair from the condensate as in (c).}
\label{fig:c2}
\end{figure}

(i)  $W$ acting on one excited pair

Adding a $\vq_3\neq\bf0$ interaction to the diagram of Fig.~\ref{fig:c1}(a) is just the same as adding a $\vq_2\neq\bf0$ interaction to the diagram of Fig.~4: we simply have to replace the $\vq_1$ scattering by a series of $(\vq_1,\vq_2)$ scatterings. As $(\vq_1,\vq_2,\vq_3)$ differ from $\bf0$, the process in Fig.~\ref{fig:c2}(a) is the only ladder process that contributes to $\Delta_N^{(3)}$.

 %It is then easy to see that the process involving two bosons, shown in Fig.~\ref{fig:5}(a), is the only one contributing to $\Delta_N^{(3)}$ because, in  a $W$ interaction,  This contribution, shown in Fig.~\ref{fig:9}(a), reads, for $\vq_2+\vq_1$ set equal to $\vq'_2$,
%\be
%\frac{N(N-1)}{2} \sum_{\vq_1\not=\bf0}\sum_{\vq'_2\not=({\bf0},-\vq_1)} \frac{v_{-\vq'_2}v_{\vq'_2-\vq_1}v_{\vq_1}}{(-2\va_{\vq'_2})(-2\va_{\vq_1}) }\label{eqc:termnn-1q1q'2}
%\ee

(ii) $W$ acting on three excited bosons

\begin{figure}[h!]
\centering
   \includegraphics[trim=2.8cm 4.3cm 4.2cm 16.6cm,clip,width=3in] {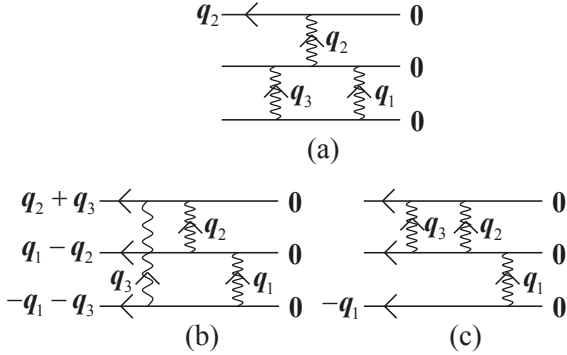}
\caption{\small The $\vq_3$ interaction takes place between two of the three bosons involved in Fig.~\ref{fig:c2}(b).}
\label{fig:c3}
\end{figure}

\begin{figure}[h!]
\centering
   \includegraphics[trim=3cm 2.3cm 3.8cm 15.2cm,clip,width=3in] {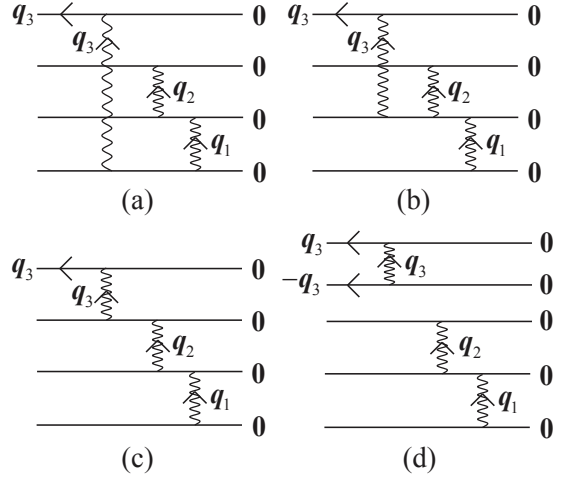}
\caption{\small The $\vq_3$ interaction involves one or two additional bosons from the condensate.  }
\label{fig:c4}
\end{figure}

Adding a $\vq_3\neq\bf0$ interaction to one of the two diagrams in Fig.~\ref{fig:c1}(b), which are topologically equivalent, leads to the diagrams in Figs.~(\ref{fig:c3}) and (\ref{fig:c4}).

--- In Fig.~\ref{fig:c3}, the $\vq_3$ interaction takes place between two of the three bosons already involved in Fig.~\ref{fig:c1}(b). So, these processes also appear with a $[N(N-1)/2](N-2)$ prefactor. As $(\vq_1,\vq_2,\vq_3)$ differ from $\bf0$, the processes of Fig.~\ref{fig:c3}(a,c) do not contribute to $\Delta_N^{(3)}$. By contrast, since we can take $\vq_1=\vq_2=-\vq_3$, the process of Fig.~\ref{fig:c3}(b) contributes to $\Delta_N^{(3)}$.

--- In the processes of Fig.~\ref{fig:c4}(a,b,c), the $\vq_3\neq\bf0$ interaction involves a fourth boson taken from the condensate, while in Fig.~\ref{fig:c4}(d), this interaction excites a $(\vq_3,-\vq_3)$ boson pair. So, the former processes appear with a prefactor $N(N-1)(N-2)(N-3)$, while the latter process appears with a prefactor $N(N-1)(N-2)(N-3)(N-4)$. However, since $\vq_3\neq\bf0$, all these processes do not contribute to $\Delta_N^{(3)}$.

\begin{figure}[h!]
\centering
   \includegraphics[trim=3cm 2.3cm 3.8cm 15cm,clip,width=3in] {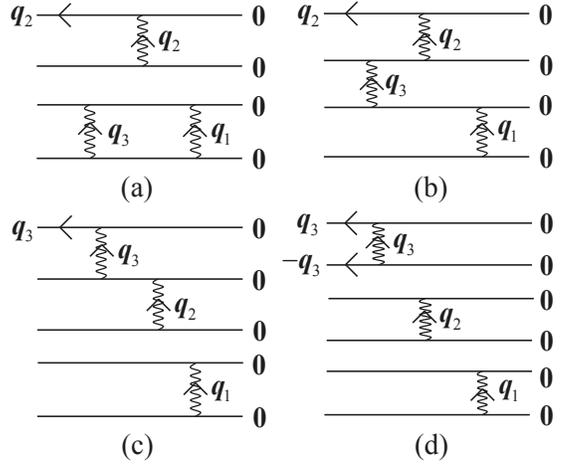}
\caption{\small Third-order interactions involving four or more bosons.}
\label{fig:c5}
\end{figure}

(iii) $W$ acting on two excited pairs

Finally, we can also add one $\vq_3\neq\bf0$ interaction to the two excited pairs shown in Fig.~\ref{fig:c1}(c).\

--- This can be done inside a pair, as in Fig.~\ref{fig:c5}(a), or between two pairs, as in Fig.~\ref{fig:c5}(b). However, since $\vq_2\neq\bf0$, these processes do not contribute to $\Delta_N^{(3)}$.

--- The third $W$ interaction can also involve one more boson from the condensate, as in Fig.~\ref{fig:c5}(c), or involve two more bosons, as in Fig.~\ref{fig:c5}(d). However, these processes do not contribute to $\Delta_N^{(3)}$ in the same way, because $(\vq_1,\vq_2,\vq_3)$ differ from $\bf0$.

\renewcommand{\theequation}{\mbox{D.\arabic{equation}}} %\section{Appendix}
\setcounter{equation}{0} %

\mbox{}\\
{\bf Appendix D. Fourth-order term in $W$}

\mbox{}

The fourth-order term in $W$ is obtained by adding a $\vq_4\neq\bf0$ interaction to the third-order state $W^3|0_N\ran$, shown in Appendix C. Processes giving a non-zero contribution to $\lan 0_N|W^4|0_N\ran$ involve two, three or four bosons $\bf0$, while those involving five bosons reduce to zero, because $W$ only contains momentum transfers that differ from zero.

\subsubsection{Two bosons involved}
We can add a $\vq_4\neq\bf0$ interaction to the ladder process in Fig.~6. For $\vq_4=-\vq_1-\vq_2-\vq_3$, this process gives a non-zero contribution in the $|0_N\ran$ condensate, as shown in Fig.~1. By setting $\vq_2=\vq'_2-\vq_1$ and $\vq_3=\vq'_3-\vq'_2$, its contribution to $\mathcal{E}_N^{(4)}=\Delta_N^{(4)}$ reads as
\bea
\lefteqn{\frac{N(N-1)}{2} \sum_{\vq_1\not=\bf0}\sum_{\vq'_2\not=({\bf0},\vq_1)} \sum_{\vq'_3\not=({\bf0},\vq'_2)}\frac{v_{-\vq'_3}v_{\vq'_3-\vq'_2}v_{\vq'_2-\vq_1}v_{\vq_1}}{(-2\va_{\vq'_3})(-2\va_{\vq'_2})(-2\va_{\vq_1}) }}\hspace{8.5cm}\nn\\
\propto Nn \, .\hspace{6cm}\ \label{eq:termnn-1q1q'2q'3}
\eea
This term contains four $v_\vq$'s, which bring four $1/L^3$ factors, three $\vq$ sums, which bring three $L^3$ factors, and a $N(N-1)/2$ prefactor; so it leads to a contribution in $Nn$. Together with the term shown in Fig.~6, this part of $\mathcal{E}_N^{(4)}$ participates in the renormalization of the $v_\vq$ interaction.
\begin{figure}[h!]
\centering
   \includegraphics[trim=3cm 6cm 2.5cm 18cm,clip,width=3.2in] {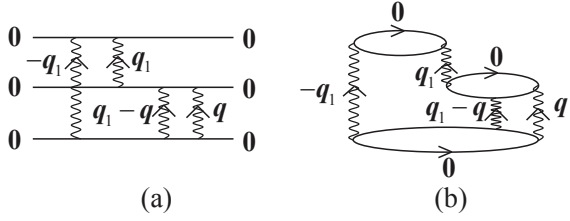}
\caption{\small This process, when combined with the three-bubble diagram in Fig.~2(a), participates in the renormalization of the $v_\vq$ interaction between bubbles through the ladder processes.}
\label{fig:7}
\end{figure}
\begin{figure}[h!]
\centering
\subfigure[]{\label{fig:8a}\includegraphics[trim=5.5cm 6cm 4cm 19cm,clip,width=2.7in] {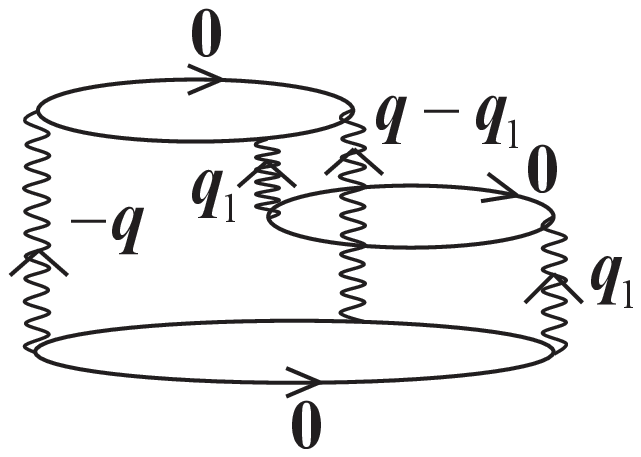}}\\
\subfigure[]{\label{fig:8b}\includegraphics[trim=3cm 5.7cm 3.5cm 18.3cm,clip,width=3.2in] {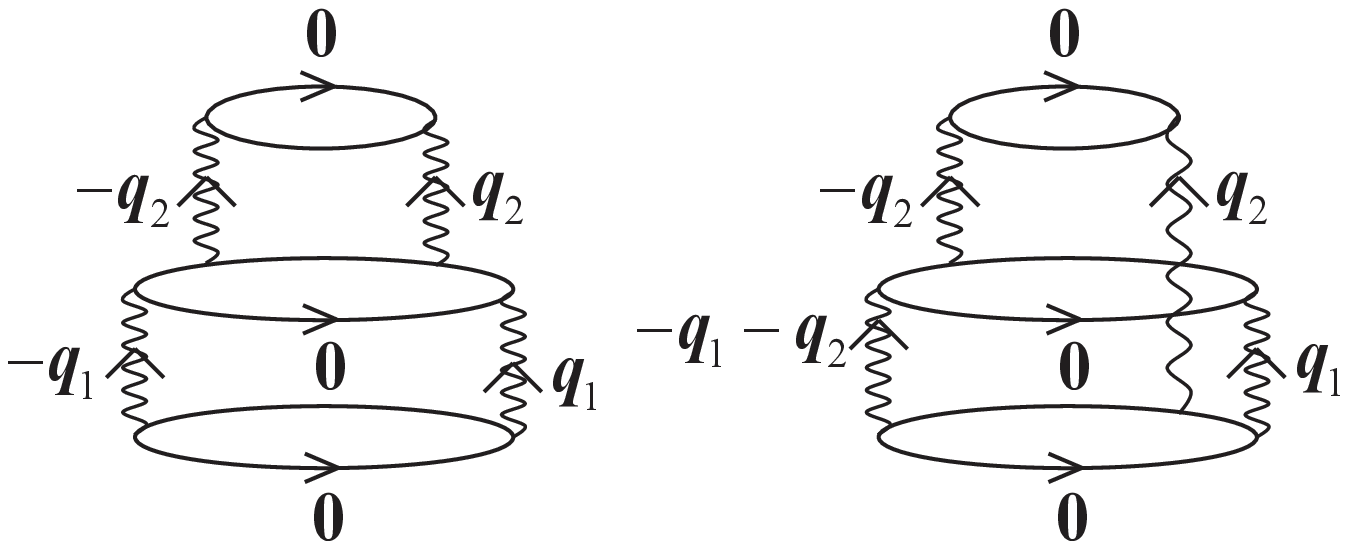}}
\caption{\small More complicated four-interaction processes between three bubbles. }
\label{fig:8}
\end{figure}
 \subsubsection{Three bosons involved}
We can add one interaction to the three-bubble processes shown in Fig.~2 by repeating one of the interactions, as in Fig.~\ref{fig:7}. So, it participates in the renormalization of the $v_\vq$ interaction between bubbles. This term contains four $v_\vq$'s, two $\vq$ sums, and a $N(N-1)(N-2)$ prefactor; so, it leads to a contribution in $Nn^2$.

We can also have more complicated three-bubble processes, in which the added interaction links different bubbles, as shown in Fig.~\ref{fig:8}(a). Finally, we can add two interactions and one bubble to the two-bubble diagram of Fig.~5, as shown in Fig.~\ref{fig:8}(b). The processes shown in Fig.~\ref{fig:8}(a) and \ref{fig:8}(b) actually contribute to higher-order density terms\cite{Wu1959,Hugenholtz1959}.

\subsubsection{Four bosons involved}

\begin{figure}[h!]
\centering
  \includegraphics[trim=1cm 2.4cm 1cm 16cm,clip,width=3.6in] {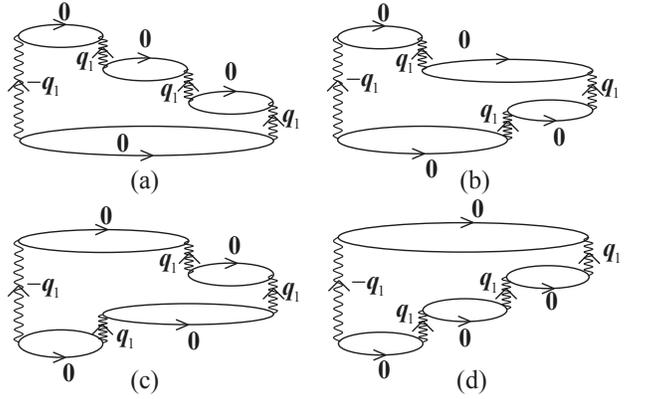}
\caption{\small  Four-bubble diagrams: the system stays with one excited boson pair $(\vq_1,-\vq_1)$ across all the intermediate steps.}
\label{fig:10}
\end{figure}

With four $W$ interactions involving four bosons $\bf0$, we can have the four-bubble diagram as shown in Fig.~\ref{fig:10}.
\begin{figure}[h!]
\centering
  \includegraphics[trim=1.5cm 1.8cm 1.5cm 15.5cm,clip,width=3.4in] {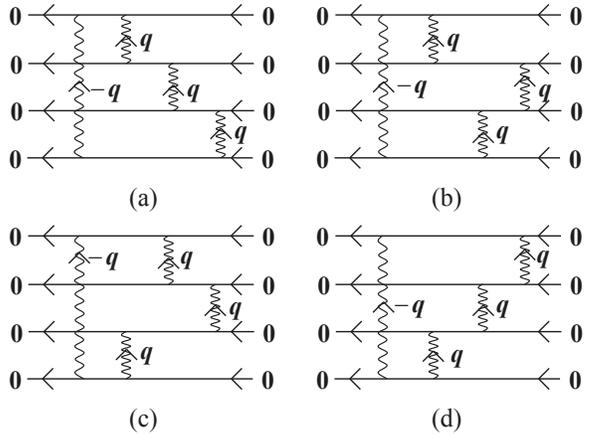}
\caption{\small  Topologically different diagrams corresponding to the four-bubble diagrams in Fig.~\ref{fig:10}.}
\label{fig:11}
\end{figure}
These bubble processes produce a contribution to $\mathcal{E}^{(4)}$, which reads as
\be
4\frac{N(N-1)}{2}(N-2)(N-3)\sum_{\vq_1\not=\bf0} \frac{v_{-\vq_1}v_{\vq_1}^3}{(-2\va_{\vq_1})^3 }\propto Nn^2\, ,\label{eq:4bosons3bubbles}
\ee
the extra $4$ factor coming from the four topologically different processes shown in Fig.~\ref{fig:11}.

\begin{figure}[h!]
\centering
   \includegraphics[trim=2.8cm 2.5cm 2.3cm 16cm,clip,width=3.3in] {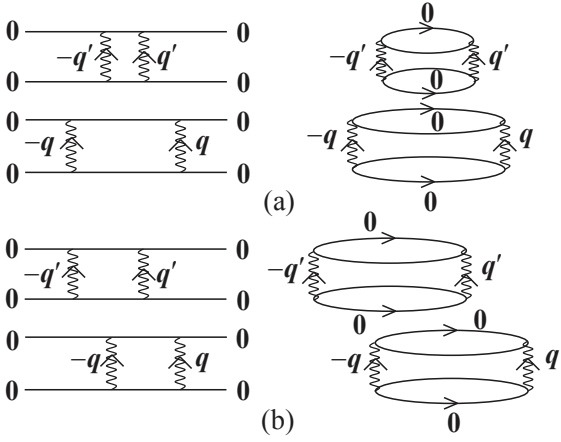}
\caption{\small  Disconnected processes resulting from four $W$ interactions between four bosons $\bf0$ from the condensate.}
\label{fig:12}
\end{figure}
We can also have four bosons involved through the disconnected processes shown in Fig.~\ref{fig:12}. They lead to
\bea
\lefteqn{\frac{N(N-1)}{2}\frac{(N-2)(N-3)}{2}}\nn\\
&&\times\bigg[\sum_{\vq\not=\bf0}\sum_{\vq'\neq\bf0} \frac{v_{-\vq}}{(-2\va_\vq) }\frac{v_{-\vq'}v_{\vq'}v_{\vq}}{(-2\va_\vq-2\va_{\vq'})(-2\va_\vq)}
\label{eq11_2nd}\\
&&+\sum_{\vq\not=\bf0}\sum_{\vq'\not=\bf0} \frac{v_{-\vq'}}{(-2\va_{\vq'})}\frac{v_{-\vq} v_{\vq'}}{(-2\va_\vq-2\va_{\vq'})}\frac{v_\vq}{(-2\va_\vq )}\bigg]\ \ \propto\ \  N^2n^2\nn\, .
\eea
These two terms contain four $v_\vq$'s, two $\vq$ sums, and a $N$ prefactor coming from the number of ways to choose the two boson pairs $(\bf0,\bf0)$ among $N$. So, they lead to an overextensive contribution in $N^2n^2$, which is canceled out by the $\Delta_N^{(2)}$ part of $J_N^{(4)}$ in Eq.~(34). Let us show it.

%\subsubsection{Five bosons involved}
%Finally, we can also have five bosons $\bf0$ involved as in Fig.~\ref{fig:10}(c,c'). This term
\renewcommand{\theequation}{\mbox{E.\arabic{equation}}} %\section{Appendix}
\setcounter{equation}{0} %

\mbox{}\\
{\bf Appendix E. Cancellation of overextensive terms}

\mbox{}

Up to now, we have shown that the $N$-boson energy at first order in interaction is equal to $v_{\bf0} N(N-1)/2$ (see Eq.~(30)), while all higher-order terms depend on $\vq\neq\bf0$ processes through the $W$ interaction. $\mathcal{E}_N^{(2)}$, given in Eq.~(42), has a unique contribution that scales as $Nn$, while $\mathcal{E}_N^{(3)}$ has a contribution, given in Eq.~(43), which also scales as $Nn$, and another contribution, given in Eq.~(44), which scales as $Nn^2$. In the same way, the part of $\mathcal{E}_N^{(4)}$ coming from the first term of $J_N^{(4)}$ in Eq.~(34) has extensive contributions in $Nn$ (Eq.~(\ref{eq:termnn-1q1q'2q'3})) and in $Nn^2$ (Eq.~(\ref{eq:4bosons3bubbles})). However, the part of $\mathcal{E}_N^{(4)}$ also has two overextensive contributions, given in Eq.~(\ref{eq11_2nd}). Actually, these two terms can be combined into
\be
\frac{N(N-1)}{2}\frac{(N-2)(N-3)}{2}\sum_{\vq\not=\bf0}\frac{v^2_\vq }{(-2\va_\vq)^2 }\sum_{\vq'\not=\bf0}\frac{v^2_{\vq'} }{(-2\va_{\vq'}) }\, .
\ee
It then becomes easy to see that, when inserted into Eq.~(34), this overextensive part is canceled out in the large $N$ limit by the $\Delta_N^{(2)}$ part of $J_N^{(4)}$.\


\begin{thebibliography}{99}
\bibitem{diagram} In this paper, we also show unconventional diagrams that are closer in appearance than Feynman diagrams to the quantities they represent. For the same reason, these ``open-line" diagrams are closer in appearance to the Shiva diagrams that we have proposed to visualize composite boson many-body effects.
\end{thebibliography}
\end{document}